\documentclass[structabstract]{aa}  
\usepackage{graphicx}
\usepackage{longtable}
\usepackage{txfonts}

\def\HeI{He\,{\sc i}}
\def\HeII{He\,{\sc ii}}

\def\HI{H\,{\sc i}}

\def\CII{C\,{\sc ii}}
\def\CIII{C\,{\sc iii}}

\def\NII{N\,{\sc ii}}
\def\NIII{N\,{\sc iii}}

\def\OII{O\,{\sc ii}}

\def\SiIV{Si\,{\sc iv}}
\def\SiIII{Si\,{\sc iii}}

\def\AlIII{Al\,{\sc iii}}

\def\Lstar{$L_{\ast}$}
\def\kms {km\,s$^{-1}$}
\def\Msunyr{\hbox{M$_\odot\,$yr$^{-1}$}}
\def\Teff{$T_{\rm eff}$}
\def\logg{$\log g$}

\def\Rstar{$R_{\ast}$}

\def\Rsun {$R_{\odot}$}
\def\Msun {${\rm M}_{\odot}$}

\def\Mstar{$M_{\ast}$}
\def\Mdot{${\dot M}$}

\def\vinf {$v_{\rm \infty}$}

\def\kms{\mbox{\rm km$\;$s$^{-1}$}}

\begin{document}
\title{A VLT/FLAMES survey for massive binaries in Westerlund 1. IV. Wd1-5 - binary product and 
a pre-supernova companion for the magnetar CXOU J1647-45?     
\thanks{Based on observations made at the European Southern Observatory, Paranal,
Chile, under programmes ESO 81.D-0324, 383.D-0633, 087.D-0440, and 087.D-0673}
}
\author{J.~S.~Clark\inst{1}
\and B.~W.~Ritchie\inst{1,2}
\and F.~Najarro\inst{3}
\and N.~Langer\inst{4}
\and I.~Negueruela\inst{5}}
\institute{
$^1$Department of Physics and Astronomy, The Open 
University, Walton Hall, Milton Keynes, MK7 6AA, United Kingdom\\
$^2$Lockheed Martin Integrated Systems, Building 1500, Langstone, Hampshire, PO9 1SA, UK\\ 
$^3$Departamento de Astrof\'{\i}sica, Centro de Astrobiolog\'{\i}a, 
(CSIC-INTA), Ctra. Torrej\'on a Ajalvir, km 4,  28850 Torrej\'on de Ardoz, 
Madrid, Spain\\
$^4$Argelander Institut f\"{u}r Astronomie, Auf den H\"{u}gel 71, Bonn, 53121, 
Germany\\
$^5$Departamento de F\'{i}sica, Ingenier\'{i}a de Sistemas y Teor\'{i}a de 
la Se\~{n}al, Universidad de Alicante, Apdo. 99,
E03080 Alicante, Spain}

   \abstract{The first soft gamma-ray repeater was discovered over three decades ago, and subsequently identified as a magnetar,
 a class of highly magnetised neutron star. It has been hypothesised that these stars 
power some of the brightest supernovae known, and that they may  form the 
central engines of some long duration gamma-ray bursts. However there is currently no consenus on the formation channel(s) of these objects.
}
{The presence of a magnetar in the starburst cluster 
Westerlund 1 implies a progenitor with a mass $\geq40M_{\odot}$, which 
favours its formation  in a binary that was disrupted at  
supernova. To test this hypothesis we conducted a search for the putative 
pre-SN  companion.} 
{This was accomplished via a radial velocity survey to identify high-velocity runaways, 
with subsequent non-LTE model atmosphere analysis of the resultant
candidate, Wd1-5.}
{Wd1-5 closely resembles the primaries in the short-period binaries, Wd1-13 and 44, suggesting a similar 
evolutionary history, although it currently appears single. It is overluminous for its 
spectroscopic mass and we find evidence of He- and N-enrichement,
 O-depletion, and critically C-enrichment, a combination of properties that is 
difficult to explain under single star evolutionary paradigms.  We infer a pre-SN history for Wd1-5
 which supposes an initial close binary comprising two stars of comparable 
($\sim 41M_{\odot}+35M_{\odot}$) masses. Efficient mass transfer from the initially more massive component leads to 
the mass-gainer 
evolving more rapidly, initiating  luminous blue variable/common envelope evolution. Reverse, wind-driven mass transfer during its subsequent  
WC Wolf-Rayet phase leads to the carbon pollution of Wd1-5, before a type Ibc supernova 
disrupts the binary system. Under the assumption of a  physical association between Wd1-5 and J1647-45,
the secondary is identified as the magnetar progenitor; its common envelope evolutionary phase
prevents spin-down of its core prior to SN and  the seed magnetic field for the magnetar forms either 
in this phase or during the earlier episode of mass transfer in  which it was spun-up.}
{Our results suggest that binarity is a key ingredient  in the formation of at least a subset of 
magnetars by preventing spin-down via core-coupling and potentially generating a seed magnetic field. The apparent
formation of a magnetar in a Type Ibc supernova is consistent with recent suggestions
that superluminous Type Ibc supernovae are powered by the rapid spin-down of these objects.}


\maketitle

\section{Introduction}

A major uncertainty in our current understanding of massive stellar evolution is the mapping of initial stellar mass onto supernova 
(SN) type and the resultant  relativistic remnant  (i.e. neutron star (NS) or black hole  (BH)). The  key driver for 
both  relationships is the magnitude of pre-SN mass loss; historically this has been assumed to be mediated by a radiatively driven 
wind but recently other modes, such as impulsive events and binary  mass transfer  have received increasing 
attention. 

Three approaches can be taken to resolve this issue (e.g. Muno \cite{muno}). First, given the association of a SN remnant 
with a relativistic object, one can attempt to  infer the properties of the progenitor object from the former; a classic example
 being the Crab nebula (Nomoto et al. \cite{nomoto}). However, this approach is model dependent, with  Cas A, for example,  being interpreted 
as originating from both single and binary progenitors of differing masses (Laming \& Hwang \cite{laming}, Young et al. 
\cite{young}). Second, theoretical reconstruction  of the evolutionary history of (high-mass) X-ray binaries such as GX301-2 (=BP Cru) 
 and 4U1700-37 (=HD 153919) from their current physical properties may be attempted (e.g. Wellstein \& Langer \cite{wellstein}, Clark et al. 
\cite{clark02}) although once again this methodology is heavily dependant on assumptions made regarding processes such as binary
 mass transfer. 

The final approach is to identify relativistic objects within their natal stellar aggregates and hence use the cluster 
properties  to infer the nature of the progenitor. This methodology is challenging. Many compact objects are ejected from 
their natal association due to SNe kicks; the host association must be demonstrated to be co-eval and in general the 
properties of the cluster population must be determined via comparison to  stellar evolutionary models.
Nevertheless, this procedure has been successfully implemented for  three clusters, each hosting a magnetar. The
brief lifetime inferred for such objects (e.g. $\leq10^4$~yr; Kouveliotou et al. \cite{kouv}, Woods \& Thompson \cite{woods}) 
implies they should still be associated with their birthsite, hence minimising false coincidences and 
allowing us to infer that their progenitor was derived from the subset of the most massive stars {\em currently} present. 
Moreover, this  also allows us to address the parallel problem of determining  the mechanism by 
which the extreme magnetic fields ($B > 10^{15}$~G; Duncan \& Thompson \cite{duncan}, Thompson \& Duncan \cite{thompson}) present 
in magnetars are generated (Sect. 5). 
The first two examples identified appear to originate from very different progenitors, with SGR1806-20 evolving from  a high-mass star
($48^{+20}_{-8} M_{\odot}$ star; Bibby et al. \cite{bibby})\footnote{The association of 1E 1048.1-5937 with a stellar wind bubble 
also points to a high-mass ($\sim30-40 M_{\odot}$)  progenitor (Gaensler et al. \cite{gaensler}).} and  SGR1900+14 from a much 
lower mass object ($\sim17\pm2 M_{\odot}$; Clark et al. \cite{clark08}, Davies et al. \cite{davies}).
However, in neither case has the cluster Main Sequence (MS) been identified, meaning that progenitor masses have been inferred from 
post-MS objects via comparison to evolutionary theory, while co-evality has also yet to be demonstrated.

The third  association is the magnetar CXO J164710.2-455216 (henceforth J1647-45; Muno et al. \cite{muno06a}) 
with the young ($\sim5$~Myr) massive ($\sim10^5 M_{\odot}$) cluster Westerlund 1 (Wd1; Clark et al. \cite{clark05})\footnote{We also highlight  the 
recent detection of a transient magnetar in the vicinity of the Galactic Centre cluster (Mori et al. \cite{mori}). However, as highlighted by these 
authors, two potential progenitor populations exist in this region, complicating the assignment of a unique  progenitor mass for SGR J1745-29.}. 
Unlike the previous
examples, its co-evality has been confirmed   from studies of both its high- and low-mass stellar cohorts (Negueruela et al. \cite{iggy10}, 
Kudryavtseva et al. \cite{ku}); thus we may safely infer the properties of the magnetar progenitor from the
current stellar population. An  absolute, {\em dynamically 
determined} lower limit to the progenitor mass of J1647-45  is provided by the $23.2^{+3.3}_{-3.0}M_{\odot} + 35.4^{+5.0}_{-4.6}M_{\odot}$
eclipsing binary Wd1-13 (Ritchie et al. \cite{ben10}). Given that the current binary period and  evolutionary states of both components of Wd1-13 
require the lower mass component to have been the initially more massive star, and adopting plausible  assumptions regarding pre-SN
 binary mass transfer (Petrovic et al. \cite{petrovic}), this rises to $\sim 40 M_{\odot}$. 
This in turn is consistent with masses inferred from the spectroscopic classification of 
the high mass component of Wd1 (e.g Clark et al. \cite{clark05}). 

Given the expected downwards revision of stellar mass loss rates due to wind clumping (Fullerton et al. \cite{fullerton}, Mokiem et al. \cite{mokiem}) the production 
of a NS from a $>40 M_{\odot}$ progenitor appears difficult. However, several theoretical studies suggest that binary driven mass loss 
can yield such an outcome, even  for  very massive ($\sim60 M_{\odot}$) progenitors (e.g. Brown et al. \cite{brown}, Fryer et al. \cite{fryer}, Yoon et al. 
\cite{yoon}). While no stellar counterpart is visible at the location of the magnetar (Muno et al. \cite{muno06a}) such an absence 
could plausibly be explained by the disruption of a putative  binary at SN. If this  were  correct, one would expect the companion to have a lower velocity  than the magnetar and hence also remain within Wd1.
Therefore, the presence  of the pre-SN companion is a clear observational 
prediction of this hypothesis and in  this paper we describe 
the  identification and analysis of a potential candidate.

\section{Data reduction and presentation}

\begin{figure}
\includegraphics[angle=0,width=9cm]{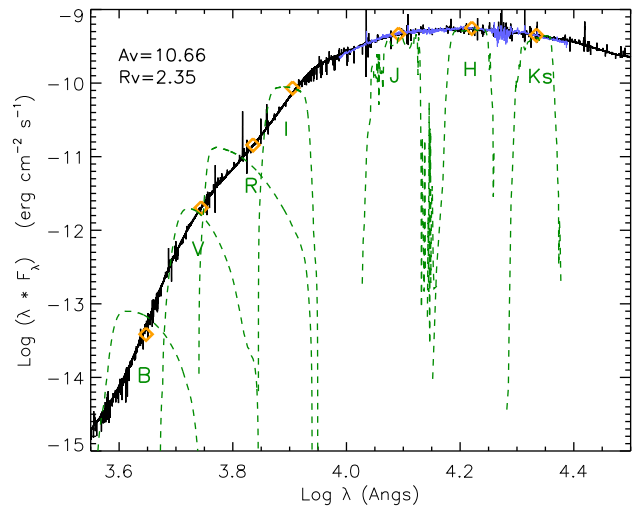}\\
\caption{Comparison of the observed and synthetic spectral energy distributions of Wd1-5. Broadband photometry is given 
by the yellow diamonds, with the corresponding 
bandpasses given by the dashed green lines. The observed and synthetic 
spectra are given by the black and blue lines, respectively.}
\end{figure}

\begin{figure}
\includegraphics[angle=270, width=9cm]{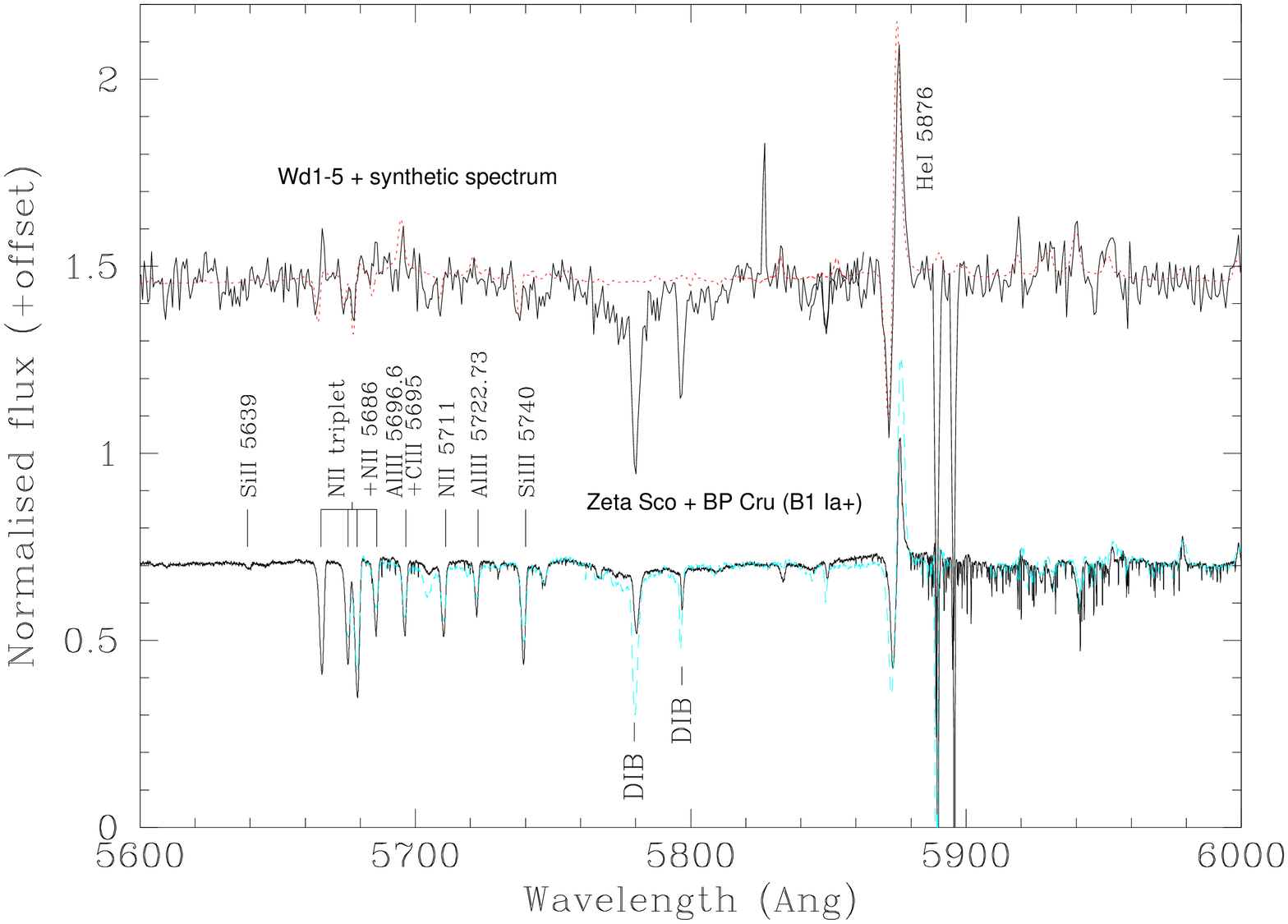}\\
\includegraphics[angle=270,width=9cm]{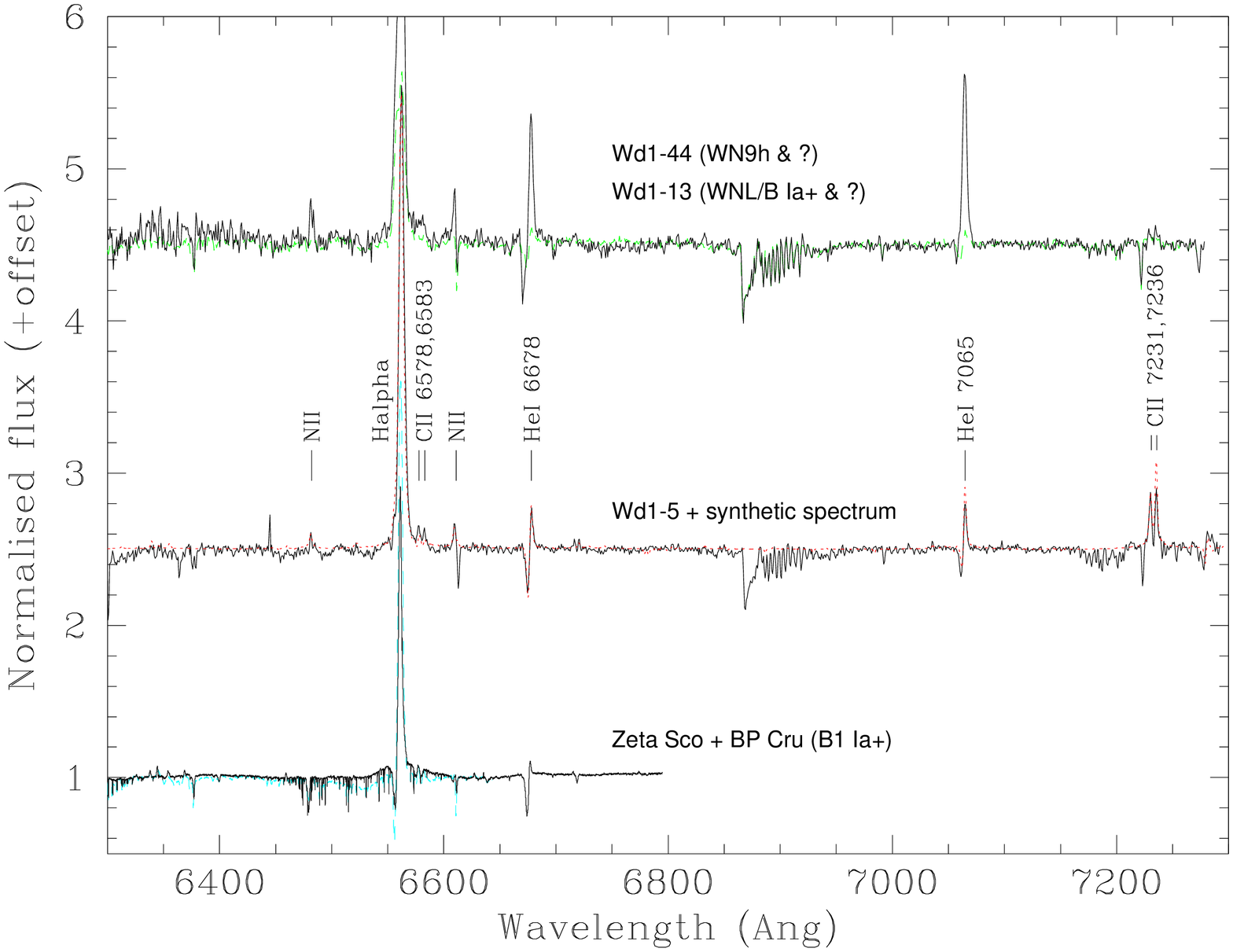}\\
\caption{ Comparison of the observed (black, solid line) to synthetic (red, dotted line) spectrum of  
Wd1-5. Illustrative spectra of selected,  closely related  early-B hypergiants and WNVLh stars are also shown 
(with BP Cru and Wd1-13 overplotted in dashed cyan and green lines, respectively,  to save space). Spectra of Wd1-13 and 44 were not available in the 
 5600-6000{\AA} window, while data on $\zeta^1$ Sco and BP Cru were from Clark et al. (\cite{clark12}) and Kaper et al. 
(\cite{kaper}). Both Wd1-13 and -44 are SB2 binaries, although a precise spectral classification of their secondaries is 
uncertain at this time.}
\end{figure}

\begin{figure}
\includegraphics[angle=270,width=9cm]{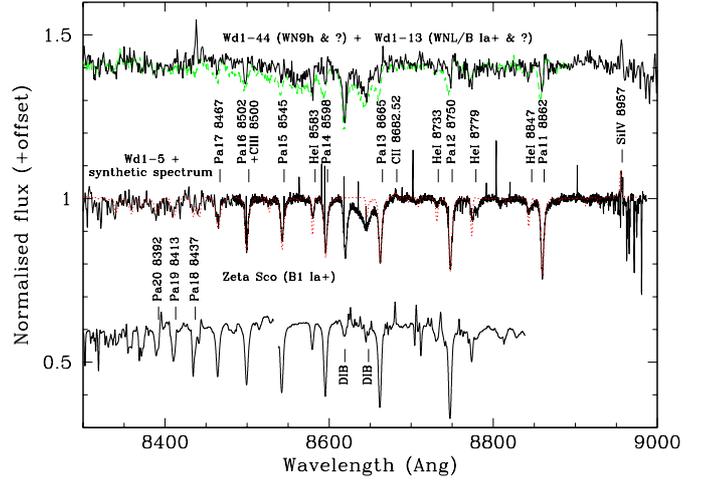}\\
\caption{Continuation of Fig. 2 encompassing the I band. Please note the difference in resolutions of the VLT/FORS2 
and FLAMES data - the latter plotted longwards of 
8475{\AA}.}
\end{figure}

\begin{figure}
\includegraphics[angle=270,width=9cm]{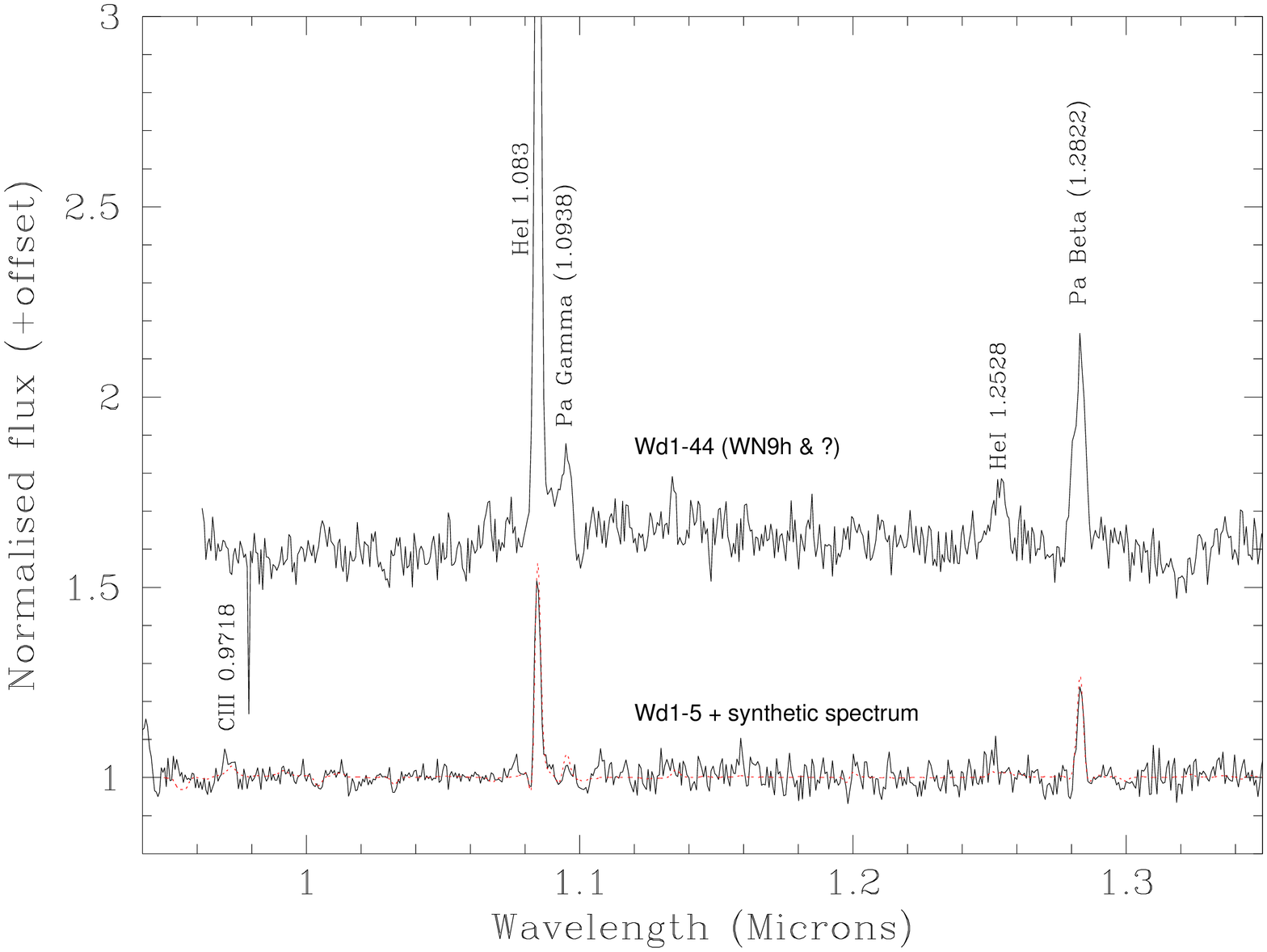}\\
\includegraphics[angle=270, width=9cm]{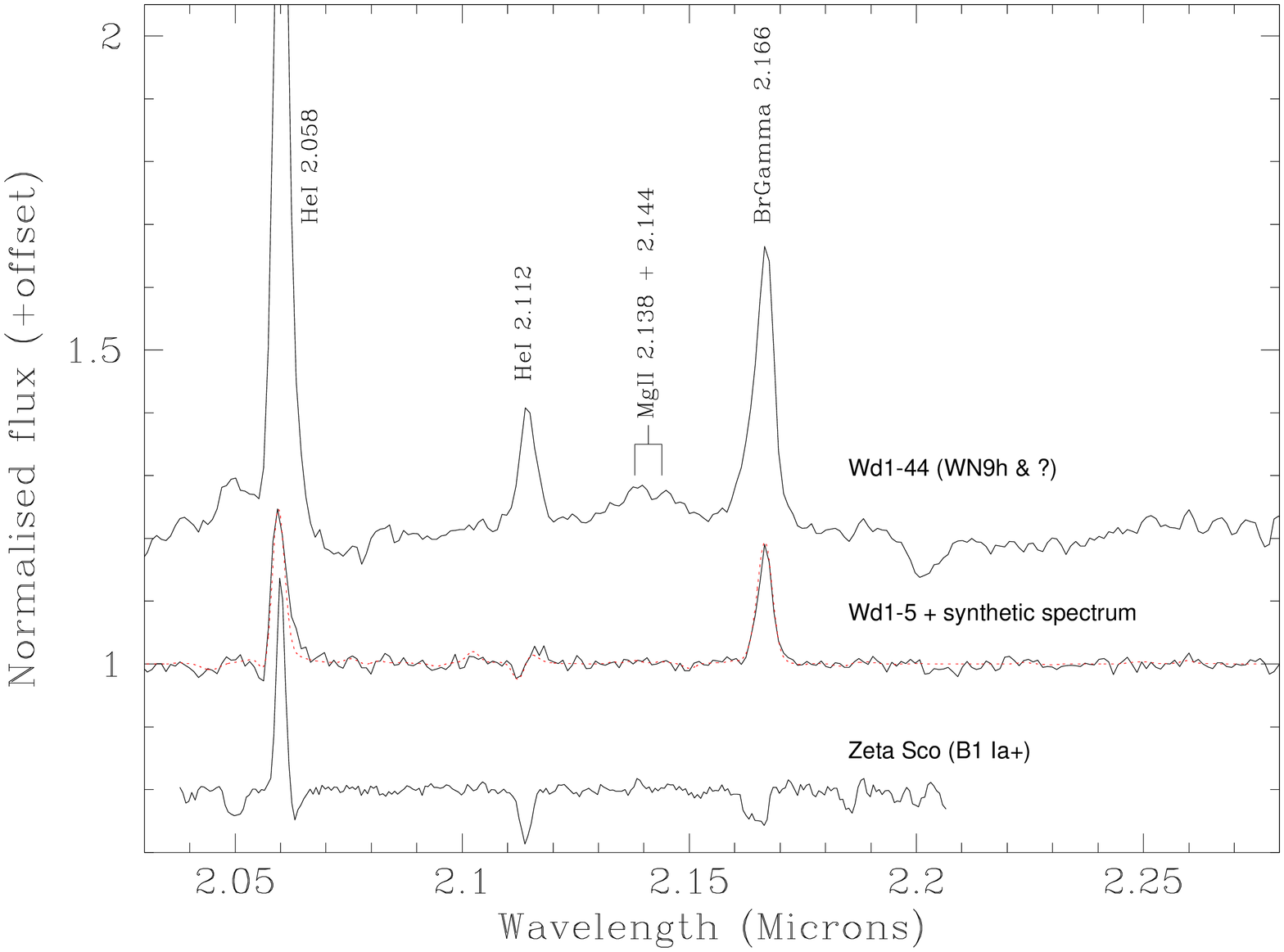}\\
\caption{Continuation of Fig. 2 encompassing the near-IR window 
(spectra from Crowther et al. \cite{crowther}); the spectra  are  of lower 
resolution than other wavebands ($\sim1000$ versus $>7000$).}
\end{figure}

\subsection{Radial velocity survey}

An obvious prediction for  a putative pre-SN companion to the magnetar is that it should have acquired an anomalous velocity 
with respect to the cluster as a result of the SN, i.e. it should be a "runaway" star (Blaauw \cite{blaauw}).

 Various authors have attempted to determine the mean systemic velocity of Wd1 via two distinct methodologies.
The first employs observations of neutral H\,{\sc i} and molecular 
material in the vicinity of Wd1 with Kothes \& Dougherty (\cite{kothes}) associating Wd1 with the
 Scutum-Crux arm ($v_{\rm sys}\sim-55\pm3$km\,s$^{-1}$) and Luna et al. (\cite{luna}) with the Norma arm  
($v_{\rm sys}\sim-90$km\,s$^{-1}$). However, by their nature they rely on the additional 
assumption of an association of kinematic and/or morphological features of the interstellar medium  with Wd~1 and 
do not directly sample the velocities of the  constituent stars (and hence do not yield a velocity dispersion). 

Three additional studies have attempted to measure the velocities of 
individual cluster members and hence the cluster systemic velocity and velocity dispersion. 
Mengel \& Tacconi-Garmann (\cite{mengel}) employ single epoch  observations of 3 red supergiants (RSGs),
5 yellow hypergiants (YHGs) and the supergiant B[e] star Wd1-9 to determine $v_{\rm sys}\sim-53.0\pm9.2$km\,s$^{-1}$, while Cottaar et al. (\cite{cottaar}) 
utilise 3 epochs of observations of the luminous blue variable  (LBV) Wd1-243 and six YHGs, of which 
5 are in common with the previous study,  to estimate a velocity dispersion of $\sim2.1^{+3.3}_{-2.1}$km\,s$^{-1}$
(but no systemic velocity determination). However, both studies are hampered by  small sample sizes, which comprise stars which are known 
pulsators and hence radial velocity (RV) variables (cf. Clark et al. 
\cite{clark10}), potentially leading to  significantly  biased RV determinations. Finally 
 Koumpia \& Bonanos (\cite{bonanos}) make multiple observations of four eclipsing binaries within Wd1 to 
determine their orbital parameters, from which mean values of $v_{\rm sys}\sim-40\pm6$km\,s$^{-1}$ are 
found  (or 
$v_{\rm sys}\sim-45\pm14$km\,s$^{-1}$ depending on the assumptions made regarding the twin components of Wd1-13). 

Between 2008-9 we undertook a multi-epoch RV survey of Wd1, utilising ESO VLT/FLAMES (Pasquini et al. \cite{pasquini}), with the primary goal of constraining the properties of the OB star binary population, but which also permits us to provide a significantly more robust estimate of its bulk  kinematic properties
(Clark et al. in prep.). Full details of target selection and data
acquisition and reduction may be found in Ritchie et al. (\cite{ben09}).
Subsequently, in 2011 we extended this 
 to include the Wolf-Rayet (WR) population (Ritchie et al. in prep.), which also permited the observation of additional
OB stars in the spare fibres.  
 Two other configurations were employed encompassing $\sim 20$ new stars, with  observations made in service mode on  2011 
April 17, May 20 and 22 and June 24 with an identical instrumental setup to previous observations\footnote{The 
GIRAFFE spectrograph operated in MEDUSA mode (HR21 setup), yielding   a spectral resolution of $\sim$16200 between  8484-9001$\AA$.}.  

Given the breadth of the emission lines in the Wolf-Rayets, these were excluded from further analysis, as were the pulsationally-prone YHGs and the LBV Wd1-243 (Clark et al. \cite{clark10}, noting that no blue hypergiants (BHGs) or RSGs were observed). 
Of the remainder, RVs were determined via Gaussian fits to the line cores of the Paschen series (individually weighted by line strength).
Previous model atmosphere analysis of stars in such a temperature regime indicated that significant wind contamination of the Paschen series is only an issue for hypergiants (Clark et al. \cite{clark12}), giving us confidence in this approach. Unfortunately, the 
He\,{\sc i} photospheric transitions were too weak to be employed for such an analysis for 
stars of  spectral types earlier than B2 (which form the  majority of our remaining sample); we note that they are  also subject to broadening, 
which is not included in the model-atmosphere code employed here for 
such high lying lines (Hillier \& Miller \cite{hillier98},\cite{hillier99})\footnote{The missing broadening mechanisms 
are relevent to the high members of the HeI 
Paschen-like series (n$\rightarrow$3) in the I band. These members approach hydrogen-like transitions and 
their theoretical profiles are expected to depart from a pure Doppler profile towards a more Stark-broadened one. 
Moreover, our current He\,{\sc i} atom splits the L$\leq3$ states only up to n=7, and packs all L states into a single 
Singlet or Triplet term for n$>7$. Thus, populations of the high lying HeI levels involved in the I band may not be
accurately accounted for in our models.}.

This analysis permitted the determination of the systemic velocity of individual objects and hence 
the cluster as a whole. In order to  accomplish this, newly identified  pulsators (in addition to the hypergiants) 
were excluded from further analysis, as 
were candidate binaries detected via reflex motion but for which  orbital solutions could not be determined 
and those stars with 3 or fewer  epochs of observations. This left a total of 61 stars from which 
we find $v_{\rm sys}\sim-42.5\pm4.6$km\,s$^{-1}$; if we also exclude {\em all} stars which
show any indication of binarity in terms of RV variability, spectral morphology,  X-ray and/or radio 
properties (Clark et al. \cite{clark08}, Dougherty et al. \cite{dougherty}, Ritchie et al. \cite{ben12}, 
Clark et al. \cite{clark13}) we are left with a reduced subset of 39 stars that yields $v_{\rm sys}\sim-42.9\pm4.6$km\,s$^{-1}$. We regard the velocity dispersion as an {\em upper limit} given that our limited timebase of observations will not be sensitive to long period binaries. 

The unusual emission line BHG Wd1-5  was identified as having  a highly discrepant 
systemic velocity - RV$\sim -99.8{\pm}1.3$km\,s$^{-1}$, or ${\sim}-56.9$km\,s$^{-1}$ 
relative to the cluster mean - suggestive of a runaway nature. No epoch to epoch 
RV shifts indicative of reflex binary motion were observed. Informed by the results of
quantitative analysis of Wd1-5, we return to a discussion of its runaway 
nature in Sect. 3 and 5.

\subsection{Dedicated observations}

Upon the identification of Wd1-5 as a potential runaway we collated existing data to constrain its nature.  Photometric data were taken from Clark et al. 
(\cite{clark05}) and Crowther et al. (\cite{crowther}) and are reproduced in Fig. 1.
Our primary spectroscopic resource were  observations made on  2011 
May 21 with VLT/FORS2  (Appenzeler et al. \cite{appenzeller}). Grisms 1028z ($7730-9480$~{\AA}), 
1200R ($5750-7310$~{\AA}) and 1400V ($4560-5860$~{\AA}) were employed with exposure times of  
$2{\times}60$s, $2{\times}600$s and $2{\times}980$s respectively. The longslit mode with a 0.3'' slit was used for all observations, 
yielding a resolution of $\sim 7000$. Data reduction  was accomplished following the  methodology described in  Negueruela et al. 
(\cite{iggy10}). The resultant spectra from both this, our higher resolution VLT/FLAMES run (Ritchie et al. \cite{ben09}) and published $JHK$-band observations (Crowther et al. \cite{crowther}) are presented in 
Figs. 2-4; we note that  low S/N as a result of 
interstellar reddening precludes meaningful discussion of the spectrum shortwards of ${\sim}5000${~\AA}.

Wd1-5 shares a similar spectral morphology to Wd1-13 and -44 (Figs. 2-4) and consequently  was initially 
assigned a 
`hybrid' classification of early-B hypergiant/WNVLh (e.g. Clark et al. \cite{clark05}). Negueruela at al. 
(\cite{iggy10})  subsequently amended this to B0.5 Ia$^+$, primarily on the basis of the similarity of the I-band 
spectrum to those of the B0.5 supergiants within the cluster; the hypergiant classification being allocated due 
to 
the presence of spectroscopic signatures of a high mass loss rate (e.g. strong H$\alpha$ emission). Note, 
however, that it is significantly fainter than the cooler (B5-9) hypergiants such as Wd1-33, although comparable 
to other OB supergiants within Wd1 (Fig. 5). Bonanos  (\cite{bonanos}) 
found  Wd1-5 to be an aperiodic photometric variable over short timescales but we find  no evidence of secular photometric or 
spectroscopic  changes over the past decade (Clark et al. \cite{clark10}). At other wavelengths Wd1-5 does not appear to 
support a near-IR excess (characteristic of colliding wind binaries containing a WC star; Crowther et al. \cite{crowther}), has a marginal X-ray detection (Clark
 et al. \cite{clark08}) and may be associated with a weak, apparently thermal radio source (Dougherty et al. \cite{dougherty}). Thus we find no 
observational
evidence for a binary companion via either direct or indirect diagnostics. In contrast, of the spectroscopically similar stars, 
 Wd-13 is a short-period eclipsing binary (Ritchie et al. 
\cite{ben10}) while recent  spectroscopy of Wd1-44 suggests a similar conclusion (Ritchie et al. \cite{ben12}) 
and  both are rather hard and bright X-ray sources, 
presumbly due to the presence of shocks in wind collision zones. 

Finally we address whether Wd1-5 could be a  chance superposition with 
Wd1. Foreshadowing the following sections,
 the reddening of Wd1-5 is fully consistent
with the mean cluster value (Negueruela et al. \cite{iggy10}). Moreover, both the 8620{\AA} Diffuse Interstellar Band and the 
Phillips (2-0) C2 band lines overlapping 
Pa-12 are identical to other B supergiants in the cluster; furthermore  the C2 lines in Wd1-5 also have radial
 velocities comparable to the cluster mean  
suggesting the material responsible for the absorption is the same in both cases. The dereddened magnitudes of 
Wd1-5, the spectroscopically similar stars Wd1-13 and -44 and the  OB supergiant population
within Wd1 are likewise directly comparable. Indeed, the sole example of 
an interloper to have been identified, 
the  O9 Iab star HD~151018 $\sim2.4$~arcmin to south of the nominal cluster centre, is easily distinguished by 
its discrepant colours and magnitude.  Regarding kinematic evidence, we note that under the assumption that Wd1-5 is an interloping field star its radial velocity would 
place it in an inter-arm void between the Scutum-Crux and Norma arms (Kothes \& Dougherty \cite{kothes}); a less intuitive scenario than the assumption it is a runaway from Wd1.
Finally,  our quantitative modelling of Wd1-5 (Sect. 3) 
indicates an anomalous pattern of chemical abudances, which is  only replicated in two other  
Galactic stars and that argues for a specific binary evolutionary pathway consistent with cluster membership (Sect. 4). 
Given the rarity of such 
objects, the similarity of Wd1-5 to other cluster members in terms of magnitude 
and reddening and the explicability of the properties of Wd1-5 in terms of binary 
evolution as a cluster member, we consider it highly unlikely that it is a distant  interloper and hence
continue under the hypothesis that it is a runaway star.

\section{Stellar properties}

\begin{figure}
\begin{center}
\resizebox{\hsize}{!}{\includegraphics[angle=0]{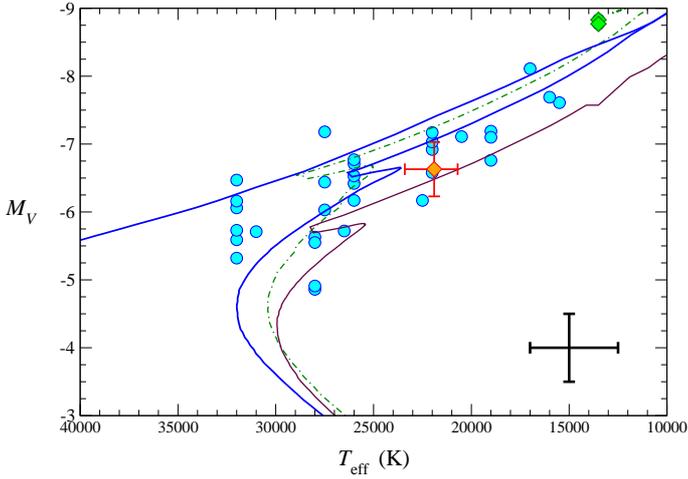}}
\caption{Semi-empirical HR diagram for Wd1-5 (diamond) and the population of bright OB supergiants  (circles) 
and hypergiants (Wd1-33 and 42; diamonds) 
 within Wd1
(following Negueruela et al. \cite{iggy10}). The solid lines represent the Geneva isochrones (Meynet \& 
Maeder \cite{meynet}) without rotation for 
log$t=6.7$ (5~Myr; {\em top}, blue) and log$t=6.8$ (6.3~Myr; {\em bottom}, brown). The dash-dotted line is the 
log$t=6.7$ isochrone for high initial rotational velocity. We note that no correction has been made for binary 
contamination in the sample, which might be expected to contribute to the scatter in absolute magnitudes.
Errorbars for Wd1-5 are those quoted in the text, while representative errorbars for the remaining stars are presented
in the lower right corner of the plot.  These represent typical uncertainties associated with an incorrect assignment of spectral type by $\pm$0.5 subtypes (in $T_{eff}$), and 0.5 mag in $M_V$ to take
into account the absolute magnitude and bolometric correction calibrations (as the
difference in $M_V$ is negligible between spectral types at Ia luminosity
class).}
\end{center}
\end{figure}

\begin{figure*}
\includegraphics[angle=0,width=14cm]{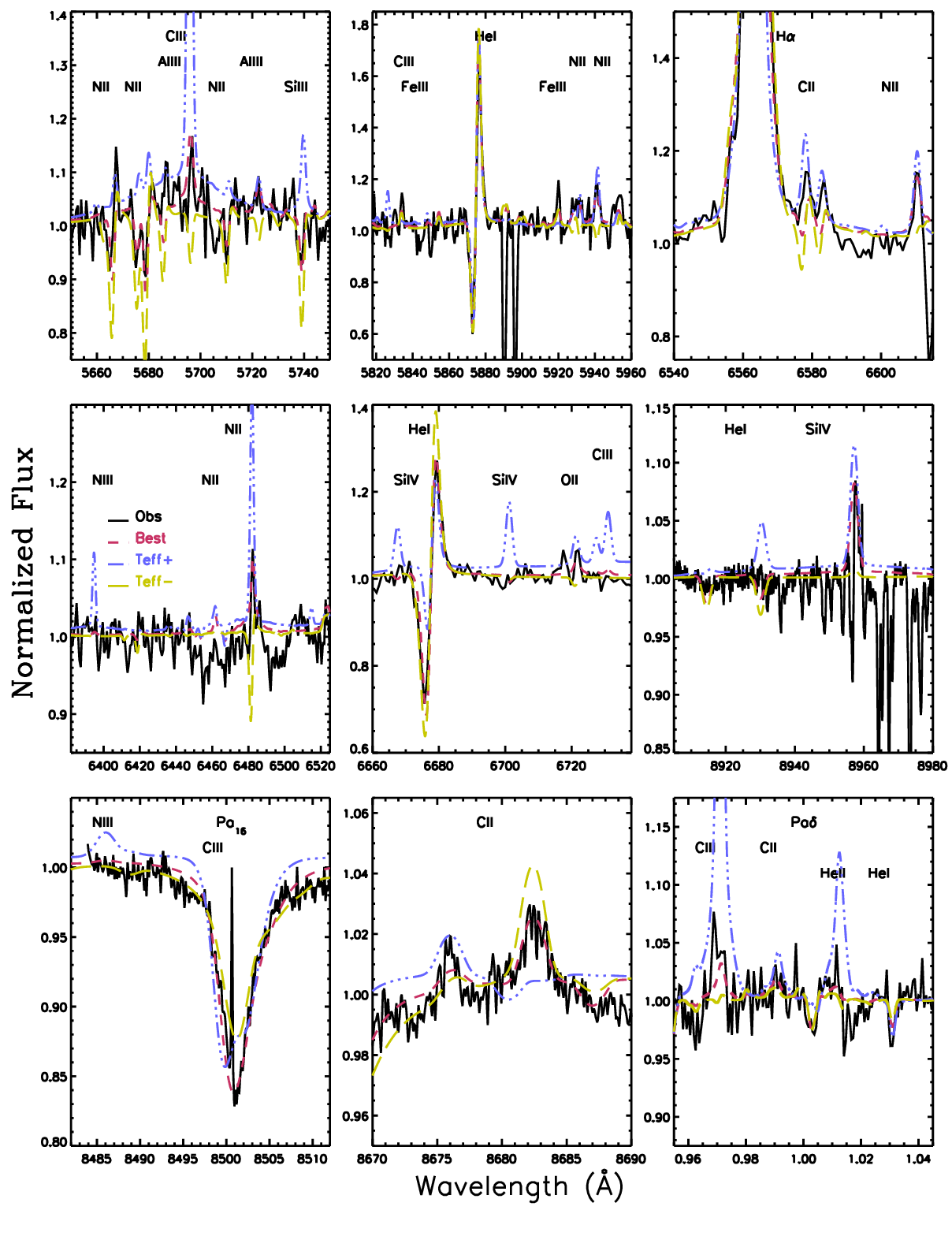}\\
\caption{Plot showing the determination of $T_{\rm eff}$ for Wd1-5. The best-fit synthetic spectrum is given
in red, with spectra utilising the upper and lower error-bounds given in Table 1 presented in blue and green 
respectively.}
\end{figure*}

\subsection{Quantitative modelling}

\begin{table*}
\begin{center}
\caption{Model parameters for Wd1-5.} 
\label{tab:model}
\begin{tabular}{ccccccccccccc}
\hline\hline
  log(\Lstar) & \Rstar     & \Teff  & \Mdot              & \vinf  & $\beta$ &  $f_{\rm cl}$  & \logg   & \Mstar & H/He & N/N$_{\odot}$ & C/C$_{\odot}$ & O/O$_{\odot}$ \\
  ($L_{\odot}$)       &  (\Rsun)     &  (kK)    &  ($10^{-6} M_{\odot}$yr$^{-1}$) & (\kms)   &         &           &  &  ($M_{\odot}$)   & &               &               &               \\
\hline
     5.38$^{+0.12}_{-0.12}$     &  34.0$^{+5.0}_{-4.4}$      & 21.9$^{+1.5}_{-1.2}$   &   2.16$^{+0.06}_{-0.07}$             &  430$^{+20}_{-40}$ &  2.50$^{+0.50}_{-0.25}$  &  0.25$^{+0.75}_{-0.15}$ &   2.33$^{+0.17}_{-0.10}$  & 9.0$^{+4.0}_{-2.0}$  & 4.0$^{+2.7}_{-1.1}$     &  9$^{+0.15}_{-0.15}$       &   1.40$^{+0.15}_{-0.15}$          &   0.3$^{+0.2}_{-0.3}$        \\
\hline
\end{tabular}
\end{center}
{We adopt a distance of $\sim5$~kpc (Negueruela et al. \cite{iggy10}) and with $A_V{\sim}10.66$ (Sect. 3) 
determine that $M_V = -6.63$ (implying a bolometric correction of -2.1~mag).
We note that \Rstar\ corresponds to $R(\tau_{\rm Ross}=2/3)$. The H/He ratio is given by number and other abundances are relative to solar 
values 
from Anders \&\ Grevesse (\cite{anders}); 
if we use the values from Asplund et al. (\cite{asplund}) as a reference, the derived ratios need to be
scaled by 1.38, 1.537 and 1.86 for C, N and O respectively. Finally, our derived H/He abundance corresponds to a surface helium mass fraction, 
$Y_{s}=0.49$. }
\end{table*}

In order to further understand the nature and evolutionary history of Wd1-5 we undertook a quantitative non-LTE 
model atmosphere analysis of it with the CMFGEN code (Hillier \& Miller \cite{hillier98}, \cite{hillier99}), utilising a 
spectroscopic and photometric dataset assembled from the above sources. Given the spectroscopic similarity to galactic BHGs such as 
$\zeta^1$ Sco and BP~Cru (Figs. 2-4) we adopted the methodology employed by Clark et al. (\cite{clark12}).

One key result of our analysis is the close proximity of Wd1-5 to the Eddington limit,
with $\Gamma{\sim}0.9$ being adopted as a {\em conservative} lower limit
(where $\Gamma_{e}{\sim}0.52$ is the contribution from electron
scattering). Moreover, since the star displays a strong wind, the stellar radius moves towards the wind/photosphere transition region,
R$(\tau=2/3)\sim 0.1 v_{\rm sound}$. These findings imply a sensitivity of the models to the \Teff~vs~\logg  ~pairings and hence  particular care was taken in 
the determination of both parameters.

\subsubsection{Temperature determination}

 To estimate the effective temperature of Wd1-5 and hence constrain the ionisation structure, we made use of 
several  ionisation equilibria, utilizing both \SiIV/\SiIII ~and \CIII/\CII\ ~line 
ratios simultaneously. The proximity of Wd1-5 to the Eddington limit, together with influence of the 
wind/photosphere transition region translates into a relatively large uncertainty in
our derived value at  \Teff=$21900^{+1500}_{-1200}$K (cf. comparable modelling of the early BHGs Cyg OB2 \#12, $\zeta^1$ Sco and HD~190603; Clark 
et al. \cite{clark12}). 
This process is shown in detail in Fig. 6,  where we display the reaction of the key lines to changes in 
\Teff ~implied by our estimated uncertainties. In this parameter domain, we found that \SiIII~5740{\AA} is 
extremely sensitive to changes in
temperature, while the  \SiIV~8957{\AA} line places a firm lower limit. Furthermore, for the
 high \Teff ~value (\Teff=$23400$K) both the \SiIV~6667{\AA} ~and 6701{\AA} transitions should be  strongly in emission, 
but are not present in our data. 
 Similar behaviour is observed in the  \CIII/\CII\ ~line ratios (see Fig. 6). A high
temperature  makes the \CII~8683{\AA} line vanish and produces excessively  strong \CIII~5695{\AA} and 9718{\AA} emission. Likewise, 
several \CIII ~lines, not detected in our spectra, appear in emission (e.g. 5826{\AA},
6727{\AA} and  6515{\AA}). While no \NIII ~and \HeII ~lines are detected,
 \NIII/\NII\ ~and \HeII/\HeI\ ~equilibria may be used as well (see Fig. 6). Many \NII ~lines
react strongly to changes in \Teff, ~while several \NIII ~lines (not detected) appear in emission for the 
 high temperature models (e.g. 6395{\AA}, 8485{\AA} and 8572{\AA}). The  \HeII~10124{\AA}  diagnostic
line suffers from poor S/N ratio and spectral resolution; however, from 
Fig. 6, the high \Teff ~value is clearly excluded. 

\subsubsection{Surface gravity determination}

To estimate \logg\  we made use of the Paschen lines in the high-resolution and signal to noise (S/N) $I$-band spectrum, as they provide the best 
constraints for the surface gravity, especially the run of the line overlap among the  
 higher members (see  Clark et al. \cite{clark12}; Fig. 3). The high Paschen lines indeed
provide reliable estimates of the stellar surface gravity, as is apparent from Fig. 7, where our current best
model together with the upper and lower logg values -  \logg=$2.33^{+0.17}_{-0.10}$
 - are compared with the observations.  We highlight  that low gravity values push  the star to the Eddington 
limit and  modify the upper photospheric and transition regions with considerable impact on the lines forming 
there (lower panels of Fig. 7). Specifically, the lack of emission in N\,{\sc ii}, Al\,{\sc iii}, Si\,{\sc iii} and He\,{\sc ii}
all exclude a lower surface gravity.

\subsubsection{Radius and mass determination}

Once the effective temperature and gravity were obtained, and assuming a distance
of $d \sim 5.0kpc$ to Westerlund~1, 
we proceeded to fit the observed optical and near-IR spectral energy distribution (SED) 
of Wd1-5 (see Fig. 1) and
hence derived the reddening, stellar radius and, therefore, the
stellar luminosity.
We used the extinction law from Cardelli (\cite{cardelli}). Several tests and comparisons with other laws were
 carried out, and very good agreement was also found with the latest extinction law from CHORIZOS (Maiz-Apellaniz, priv. comm.).
 We found $E(B-V)=4.54$ and a reddening parameter
$R_V=2.35$, corresponding to $A_V=10.66$. Such a finding is fully consistent
with the  mean value found for OB supergiants within Wd1 ($E(B-V)=4.2 \pm 
0.4$ with a $1-{\sigma}$ 
standard deviation; Negueruela et al. \cite{iggy10}).

Fig. 1 also displays
the excellent agreement between our model and the near-IR flux calibrated spectra obtained with
SOFI,
together with the filters used in the photometry (green-dashed lines).
From these values we obtained a stellar radius of  $34^{+5.0}_{-4.4}R_{\odot}$, corresponding to a 
spectroscopic mass of M=$9^{+4}_{-2}$\Msun and a
final luminosity of  $2.39^{+0.78}_{-0.56}\times 10^5 L_{\odot}$. We note that the error in the luminosity is dominated
by the uncertainty in the reddening determination. Finally, while we strongly favour a distance to Wd1 of 
$\geq 5 kpc$ (Negueruela et al. \cite{iggy10}), a value of $4 kpc$ (the lower end of literature values) would lead to a revision in radius by a factor of 
$(d/5{kpc})\sim0.8$, luminosity by $(d/5{kpc})^{1.72} \sim0.68$ and mass by $(d/5{kpc})^{2.0}\sim0.64$. 

\subsubsection{Determination of wind properties}

Although the moderate reddening affecting Wd1-5 prevents
us from securing UV observations from which we might derive firm
\vinf\ estimates, the dense stellar wind provides
alternative \vinf\ diagnostic lines such as H$\alpha$, 
\HeI~10830{\AA} and \HeI~20581{\AA}. Since the latter two were observed at relatively  low ($\sim$1000)
resolution and moderate ($\sim50$) S/N, we basically rely on  H$\alpha$ as primary diagnostic to derive
\vinf\ and $\beta$, the parameter controlling
the shape of the velocity field. 
We obtain \vinf=$430^{+20}_{-40}$km\,s$^{-1}$ and $\beta=2.5^{+0.5}_{-0.25}$. The uncertainties are mainly driven
by the moderate S/N at H$\alpha$  due to the reddening.

The main observational constraints which set the mass-loss rate and clumping are the optical H$\alpha$,
\HeI~5875{\AA}, \HeI~6678{\AA} and \HeI~7065{\AA} lines  and the near-IR  Pa$\beta$,  Br$\gamma$,  \HeI~10830{\AA} and
\HeI~20581{\AA} emission lines. As shown by Najarro et al. (\cite{paco97}, 
\cite{paco06}) the latter
is extremely sensitive to modelling assumptions and atomic data and is less reliable as an 
\Mdot ~diagnostic; nevertheless, we are still able to reproduce this feature in our synthetic spectrum (Fig. 4).
Thus, we were able to derive a mass-loss rate of \.{M}=$2.16^{+0.06}_{-0.07}
 \times10^{-6} M_{\odot}$yr$^{-1}$~(see Table~1) and a clumping factor of 
$f_{\rm cl}=0.25^{+0.75}_{-0.15}$. The latter is essentially
set by the electron scattering wings of H$\alpha$ and its error estimates
are dominated by the 
uncertainties  in \Teff\ and \logg\ together 
with the error in the normalisation of the  spectra as a result of the S/N in the
line. Note that the mass loss scales as ($d/5$kpc)$^{1.43}$.
We may compare  the unclumped 
\.{M}$=(2.16/0.25^{0.5})\times 10^{-6}M_{\odot}$yr$^{-1}=4.32\times 
10^{-6}M_{\odot}$yr$^{-1}$ to the theoretically
predicted one (Vink et al. \cite{vink00}). For Wd1-5 stellar parameters
we obtain \Mdot=$3.0 \times 10^{-5}$\Msunyr, a factor of 7 above our value. However, if we just increase
our derived mass from 9 to 10.5\Msun~(a change of 0.07 in \logg), the theoretically predicted \Mdot\ drops
to \Mdot=$7.1 \times 10^{-6}$\Msunyr, (i.e. just a factor of 1.6 above our value). Thus, we  conclude
that Wd1-5 may lie in the bi-stability jump region.

\subsubsection{Abundance determinations}

As the derived abundances play a crucial role in our conclusions regarding the nature of
Wd1-5, we describe their determination here. In all cases the uncertainties in the
abundances take into account the uncertainties
in \Teff\ and \logg\ - i.e. they are estimated considering possible combinations
of \Teff\ and \logg\ within the accepted range.
The helium abundance is set by the relative strength of the
\HI\ and \HeI\ emission lines. We derive He/H=0.25 by number (49\% by mass) with
0.15 and 0.35 as lower and upper limits.

A relatively enhanced carbon abundance ($\sim 1.4 \times$ solar) is obtained from the
\CII~6578-6583{\AA}, \CII~7231-7236{\AA} and \CII~8683-8697{\AA} lines.  The \CIII~9717{\AA} ~and the \CII~9904{\AA}
~lines present in the low resolution noisy J band spectrum are used as a consistency  check. Fig. 8
illustrates the sensitivity of the available carbon lines to changes in the abundance.  As with Figs. 6 \& 7, 
models labeled as C+/- correspond to changes in $\pm$0.15~dex (our estimated uncertainty), while those labeled as
 C++/- - have variations of $\pm$0.35~dex in the carbon abundance.
The \CIII\ line at 5695.9{\AA} is blended with the \AlIII~5696.6{\AA} transition, while the
\CIII~8500{\AA} line is located within the blue absorption wing of Pa-16, and therefore both are used as  
secondary abundance diagnostics. The same weighting is given to the
\CII~6578-6583{\AA} lines in the red wing of H$\alpha$, as they are polluted by the electron
scattering  emission wing and to \CII~8696.7{\AA}, as it is blended with \NII~8697.8{\AA}. Following from Fig. 8  
we therefore estimate  an uncertainty of $\sim 0.15$dex for our derived carbon abundance, confirming the abnormally
high value for the present evolutionary stage of the object. We return to this in Sect. 3.2.

The nitrogen abundance is basically constrained from the large number of \NII\ lines throughout the
full spectral range covered by our observations. We find a significant enhancement 
($\sim 9\times$ solar) consistent with moderate CNO processing. Again a value of 0.15dex should be 
regarded as a safe estimate for the error in the N abundance.
Finally, to derive the oxygen abundance we make use of the weakly detected \OII\ emission lines
at 6641 and 6721{\AA} and the  \OII\ absorption lines at 8564 and 8687{\AA} and obtain a value
corresponding to $\sim 0.3$ solar. In this case, the uncertainty on the derived abundance is much higher and 
we obtain 0.2 and 0.3dex respectively for the higher and lower error estimates.

\subsubsection{Additional implications}

The close proximity of Wd1-5 to the Eddington limit implies that
even moderate rotation may place Wd1-5 close to critical, or breakup velocity
 ($v_{\rm crit}$; Langer \cite{langer}, Maeder \& Meynet \cite{maeder}). 
Utilising the stellar parameters from Table 1 and assuming 
$\Gamma{\sim}0.9$  we may estimate $v_{\rm crit}\sim 80$km$^{-1}$. 
We obtain an upper limit to the  projected rotational velocity of $v_{\rm rot} \sim 60$km\,s$^{-1}$
from the  profile of the narrow C\,{\sc ii} 8683{\AA} photospheric emission line,
 which would be consistent with $v_{\rm crit}$ for 
inclinations, $i < 49^{\rm o}$; unfortunately we currently  have no constraints on the inclination of Wd1-5. In this regard we note that the 
spectroscopic mass   given in Table 1, derived  from log$g$ has consequently not been 
corrected for rotation and hence should properly be regarded as a lower limit; of importance for comparison to the stellar structure scenarios 
presented for Wd1-5 in Sect. 3.3.

Finally, an important finding from this analysis was that cross correlation of the synthetic to  the high-resolution spectrum ($\sim8484-9000${\AA}) resulted 
in a significantly lower RV shift of $\sim-68\pm4$km\,s$^{-1}$  relative to rest-wavelength when compared to  that 
determined from Gaussian line-core fitting ($\sim-99.8{\pm}1.3$km\,s$^{-1}$; Sect. 2.1). We suspect this is due to the presence of excess 
emission in the red flanks of the Paschen series photospheric lines (due to heavy mass loss), which systematically drives the line centres to shorter wavelengths. Indeed, 
cross correlation of the spectrum of Wd1-5 to that of a normal supergiant (the BI.5 Ia star Wd1-8b) and a synthetic spectrum of a supergiant of comparable 
temperature but of higher surface gravity and weaker wind resulted in an RV  offset relative to rest-wavelength 
of over $-90$km\,s$^{-1}$ in both cases.
We conclude that the true RV shift of Wd1-5 relative to the cluster mean is smaller than the initial determination 
of $\sim 56.9$km\,s$^{-1}$, but in the absence of an {\em observational}
template, its determination is  dependant on the model parameters adopted for the generation of the synthetic spectrum.
 We therefore adopt a conservative
lower limit to  the offset from the cluster systemic velocity of $>25.1$km\,s$^{-1}$; discrepant at the  
 $\sim 5.5{\sigma}$ level with the mean velocity dispersion of stars within Wd1.

Following from the initial discussion
 regarding the identification of runaways (Blaauw et al. \cite{blaauw}), various studies have refined the criterion for
 runaway status for 1-dimensional radial velocity data (Vitrichenko et al. \cite{vitrichenko}, Cruz-Gonzalez et al. \cite{cruz}, Tetzlaff et al. \cite{tetzlaff}). 
These have resulted in a downwards revision in the absolute radial velocity threshold to 25km\,s$^{-1}$ and also the adoption 
of a more generic threshold of $v > 3\sigma$, where $\sigma$ is the 1-dimensional mean velocity dispersion of low-velocity 
stars. Wd1-5 appears to satisfy both criteria, although we note that the former is defined via the motion of field stars 
within 3kpc of the Sun (Tetzlaff et al. \cite{tetzlaff}) and hence there is no {\em a priori} reason for this to match the 
mean velocity 
dispersion  of stars within Wd1.

\begin{figure}
\includegraphics[angle=0,width=9cm]{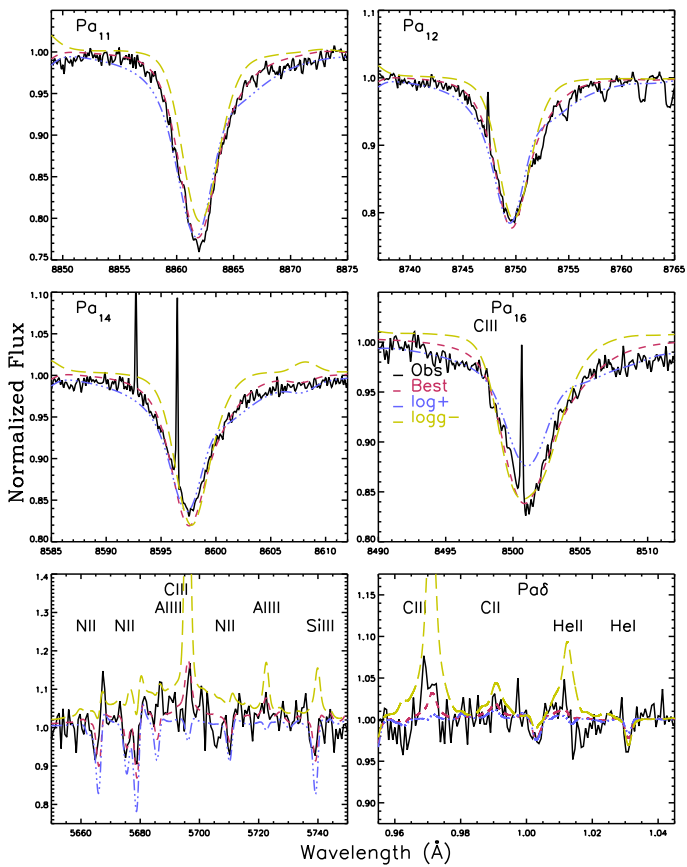}\\
\caption{Plot showing the determination of log$g$ for Wd1-5. The best-fit synthetic spectrum is given
in red, with spectra utilising the upper and lower error-bounds in Table 1 presented in blue and green 
respectively.}
\end{figure}

\begin{figure}
\includegraphics[angle=0,width=9cm]{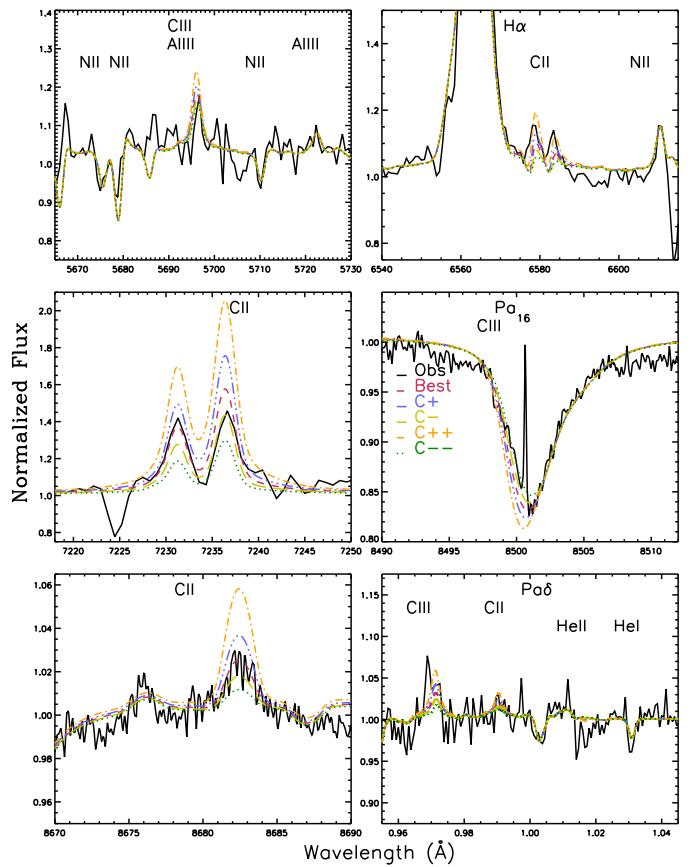}\\
\caption{Plot showing the determination of the carbon abundance for Wd1-5. The best-fit synthetic spectrum is given
in red, with spectra utilising the upper and lower error-bounds given in Table 1 presented in blue and green 
respectively. Additionally, we provide spectra  produced using more extreme, observationally unsupported upper and lower error-bounds ($\pm0.35$dex; orange and dark green respectively).}
\end{figure}

\begin{figure}
\begin{center}
\resizebox{\hsize}{!}{\includegraphics[angle=0]{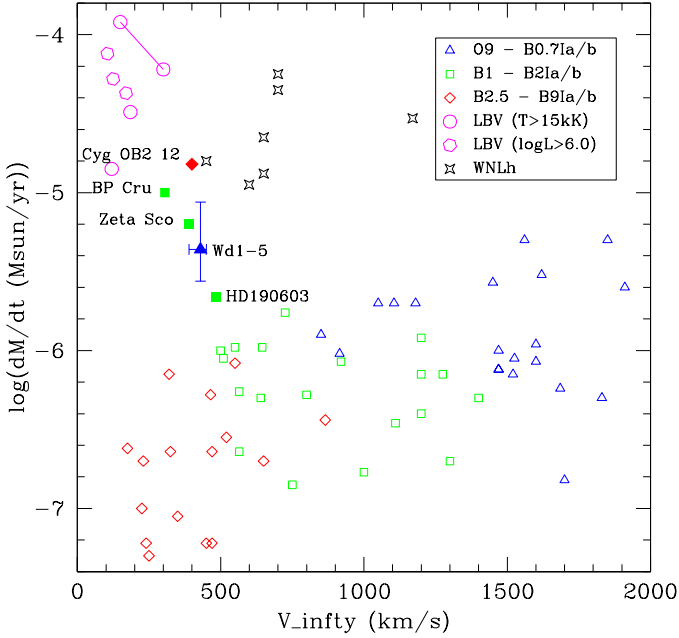}}
\caption{Comparison of the wind properties of Wd1-5 to field  BHGs, BSGs (both plotted according to spectral type, with BHGs given by solid symbols), LBVs and WNLh (=Ofpe/WNL) stars.  The data and references employed are summarised
in Table A.1. Unfortunately, given the diverse sources employed to construct this figure, it is not possible to 
plot  representative errors for all the objects.
In all cases clumping corrected mass loss rates are used to enable direct comparison between individual 
objects. We estimate  log\.{M}$\sim -5.36^{+0.3}_{-0.2}M_{\odot}$yr$^{-1}$  for Wd1-5 since the errors 
on the clumping factor in Table 1 are conservative and once this parameter is set the mass loss rate is also determined 
to a high precision. }
\end{center}
\end{figure}

\subsection{Comparison to related objects}
 In order to place Wd1-5 into an astrophysical context it is instructive to 
compare its  physical properties to those  of other galactic BSGs, BHGs, LBVs and Ofpe/WNL (=WN9-11h) stars and related 
objects. Quantitative analyses of such stars have been undertaken by  a number of authors and are summarised in 
Table A.1. The temperature and radius (and hence luminosity) of Wd1-5 are directly comparable to 
those of early spectral-type field BSGs;  conversely, both BHGs and LBVs appear significantly more physically extended than 
Wd1-5, which also lies at the high temperature extreme found for these stars. Ofpe/WNLh stars span a wide 
range of temperatures and luminosities - presumably corresponding to a spread in initial 
masses - and the temperature 
and radius of Wd1-5 appears comparable to the least luminous examples, subject to the large uncertainties in 
these parameters (Table A.1). 

Systematic quantitative analysis has yet to be undertaken for the wider population of massive evolved stars within Wd1.
Nevertheless, following Negueruela et al. (\cite{iggy10}) we may construct a semi-empirical HR diagram for Wd1 (Fig. 5),
which demonstrates the  close correspondance between Wd1-5 and the cluster OB supergiants, noting that we might expect all
these objects to be of similar mass given the co-evality of the population.

Conversely,  the wind properties of Wd1-5 do not closely resemble those of either BSGs or Ofpe/WNLh stars, despite 
similarities in temperature and luminosity. With regard to the former, the wind terminal velocity of Wd1-5 is 
significantly lower than found for BSGs of comparable spectral type (Fig. 9). In comparison to Ofpe/WNLh stars, both the terminal 
velocity and mass loss rate of Wd1-5 are at the extreme lower bounds experienced by  such stars. A similar discrepancy is found in comparison to the  LBVs, which drive slower winds with higher mass loss rates than Wd1-5 (Fig. 9). Indeed, as expected from their spectroscopic 
similarity, Wd1-5 most closely resembles other BHGs in terms of wind properties, {\em despite being significantly more physically compact than such stars} (Table A.1.). Finally, while we may not readily determine the surface gravity of LBVs and Ofpe/WNLh stars due to the lack of suitable spectroscopic diagnostics, the surface gravity of Wd1-5 appears  lower than any of the 20 O9-B0.7 Ia stars sampled in Table A.1, but once again consistent with those found for BHGs.

Therefore, in terms of the combination of both stellar (temperature, radius and surface gravity) and wind (mass loss rate, terminal velocity) properties we are unable to identify a comparable star to Wd1-5, which resembles an OB supergiant in terms of radius but a BHG in terms of surface gravity and wind properties. We return to the implications of these findings in Sect. 3.3.

Lastly  we turn to chemical abundances. Both Crowther et al. (\cite{pacBSG}) and Searle 
et al. (\cite{searle}) assume moderate H-depletion (H/He$\sim4.0$) for Galactic   BSGs, 
and find N-enrichment and C- and O-depletion, consistent with the products of CNO burning. Clark et al. (\cite{clark12}) provide a detailed 
analysis of the abundance patterns of early-B hypergiants. As with BSGs, the three stars studied - Cyg OB2 \#12, $\zeta^1$ Sco and HD~190603 - 
exhibit the products of CNO burning, with carbon being very depleted (C/C$_{\odot}\sim$ $0.21\pm0.2$, $0.33\pm0.2$ and $0.33\pm0.2$ 
respectively). This differs from  the super-solar abundances we infer for Wd1-5 (C/C$_{\odot}\sim 1.4\pm0.15$) at $>5\sigma$ level. A similar pattern is also observed for those LBVs for which similar analyses have been performed - C/C$_{\odot}\sim$ $0.11\pm0.03$, 0.31 and trace for AG Car, P Cygni and Wd1-243 respectively (Groh et al. \cite{groh}, Najarro \cite{paco01} and Ritchie et al. 
\cite{ben09b}).

 This behaviour (He and N enhancement and C and O depeletion) is also  predicted for 
single stars by current  evolutionary codes (Ekstr\"{o}m et al. \cite{ekstrom}). Stars of comparable luminosity to Wd1-5 are 
expected to encounter the BHG phase either side of a red-loop across the HR diagram; prior to this they are expected to exhibit moderately
sub-solar (C/C$_{\odot}\sim0.2-0.5$) carbon abundances  and after this passage they are extremely depleted 
(C/C$_{\odot}\sim 0.02-0.03$) due to mass-stripping as a RSG (Jose Groh, 2013, priv. comm.).

Therefore, the C-abundance of Wd1-5 is unexpected on both theoretical and observational grounds. We highlight that even in the absence 
of our quantitative analysis, the evidence for a high C-abundance is compelling; simple comparison of the optical spectra of Wd1-5 to the two other cluster BHG/WNLh stars - Wd1-13 and -44 - reveals the anomalous strength of the C\,{\sc ii} 7231{\AA} and 7236{\AA} lines in Wd1-5, with these lines being essentially absent in its spectroscopic twins (Fig. 2). All three stars are expected to share the same natal metalicity and evolutionary pathway (since Wd-1 appears essentially co-eval; Negeuruela et al. \cite{iggy10}, Kudryavtseva et 
al. \cite{ku}) and so it is difficult to account for this observational finding unless Wd1-5 has a greater C-abundance than these objects.

To date, the  sole exceptions to these abundance patterns are the B1 Ia$^+$ hypergiant BP Cru (=Wra977) and
the O6.5Iaf$^+$ star HD~153919; the mass donors in the high mass X-ray binaries GX301-2 and 4U1700-37 respectively. In addition to
 the expected  He- and N-enrichment, both show significant C-enrichment over CNO equilibrium values (C/C$_{\odot}\sim2$ and 1 respectively; Kaper et al. \cite{kaper},  Clark et al. \cite{clark02}) in an analagous manner to Wd1-5; the implications of this finding are discussed below. 

\subsection{Internal structure and evolutionary state}

Given the discrepancies in the properties of Wd1-5 when compared to other massive evolved stars 
we may also employ our quantitative analysis to try to determine the internal 
structure - and hence evolutionary state - of Wd1-5 via comparison to theoretical predictions. An immediate difficulty is encountered when 
attempting to reconcile  the high intrinsic luminosity with the spectroscopic mass estimate, 
in the sense that it appears significantly overluminous for a star of normal composition.  If Wd-1 were found 
at a lower distance than the $5kpc$ adopted here this  discrepancy becomes worse in the sense that stellar mass 
decreases more rapidly as a function  of distance than  luminosity does. Thus, while the absolute masses of the models 
described below would be reduced, our qualitative conclusions  would remain entirely unchanged.

In order to replicate 
our modelling results we investigated three different scenarios to overcome this limitation:

\paragraph{\bf Scenario (i):} Wd1-5 is chemically homogeneous. From  Eqn. 11 of Graefener et al. (\cite{graefener}) we derive   $M=32.5M_{\odot}$, leading 
   to log$g=2.79$, $\Gamma_e=0.168$, and an escape velocity, $v_{\rm esc}=520$km\,s$^{-1}$.

\paragraph{\bf Scenario (ii):}  Wd1-5 is  core-helium burning and  possesses a shallow H-envelope. From  Eqn. 18 of Langer (\cite{langer89}) we find  $M=13.7M_{\odot}$, 
  resulting in log$g=2.41$, $\Gamma_e=0.40$ and  $v_{\rm esc}=290$km\,s$^{-1}$.

\paragraph{\bf Scenario (iii):}  consisting of a hybrid of the  above, whereby Wd1-5 has burnt hydrogen in the core
   to the stage of $X_{\rm core}=0.1$. and has little envelope mass. 
Then we may again employ Eqn. 11 of Graefener et al. (\cite{graefener}) to determine  $M=18.6M_{\odot}$,
   log$g$=2.54, $\Gamma_e=0.30$, and $v_{\rm esc}=360$km\,s$^{-1}$.\newline

Scenario (i) appears excluded on the basis of pronounced discrepancies between 
the predicted and observed values of log$g$ (and hence stellar mass), wind velocity 
($v_{\rm esc}$ versus $v_{\infty}$) and $\Gamma_e$. 
However, scenarios (ii) and (iii) - that Wd1-5 is essentially an He-star with relatively shallow H-envelope - 
appear more acceptable in terms of these properties. In such a picture  $v_{\infty}$  exceeds the predicted $v_{\rm esc}$, 
as might be expected and $\Gamma_{e}$ approaches
 the value of 0.5 determined via modelling, noting 
that for Galactic metallicities, the Eddington factor based on the full Rosseland mean opacity in the atmosphere of hot 
massive stars is larger by $\sim$0.3  compared to the case when only electron scattering is considered as an opacity 
source. Likewise, the discrepancy between predicted and spectroscopic  masses (Table 1) may be ameliorated by 
the inclusion of the (uncertain) correction due to the effects of stellar rotation, given the proximity of Wd1-5 to the Eddington limit 
(Sect. 3.1). 

Nevertheless, if Wd1-5 were born with a mass at the main sequence turn-off of Wd1 ($\sim 40M_{\odot}$; 
Ritchie et al. \cite{ben10}),  we would find it difficult to account simultaneously  for the luminosity of Wd1-5 and 
the observed, He-rich surface chemistry under the assumption of mass loss driven by stellar winds (Brott et al. 
\cite{brott}). Moreover, we would not expected it to achieve such an apparently low mass while retaining such a 
high hydrogen content. Conversely, while the 
current mass would be explicable if Wd1-5 was born  with  a lower initial mass ($\sim20M_{\odot}$),
 we would not expect  such a star to have evolved from the main sequence given the current age of Wd1 (Negueruela et al. \cite{iggy10}). Furthermore,  the requirement for stellar winds to yield the observed surface chemistry would be even more 
implausible. 

As well as He-enrichment, Wd1-5 is also N- and, critically, C-enriched. 
Unlike the He- and N-enrichment, which  in principle might be understood  as the  result of products of  the CNO cycle 
being transported to the surface via rotational mixing, the C-enhancement over CNO equilibrium values {\em cannot} arise from such a mechanism, since
 any carbon produced by  He-burning would have to pass through overlying H-burning layers prior to reaching the surface where it would 
be converted into nitrogen (e.g. Clark et al. \cite{clark02}). Consequently, this cannot be the result of the evolution of a single star and must be the result of mass-transfer from a C-rich bianry companion.

{\em Therefore, in order to reconcile the luminosity, mass and surface composition, we are forced to conclude that Wd1-5 must have been subject to an additional source of mass-loss, which we suppose was binary-induced, since we also must infer (subsequent) mass-transfer from a putative binary companion.}

\section{The nature and formation of Wd1-5}

\subsection{A pre-SN binary evolutionary pathway}

\begin{figure}
\begin{center}
\resizebox{\hsize}{!}{\includegraphics[angle=0]{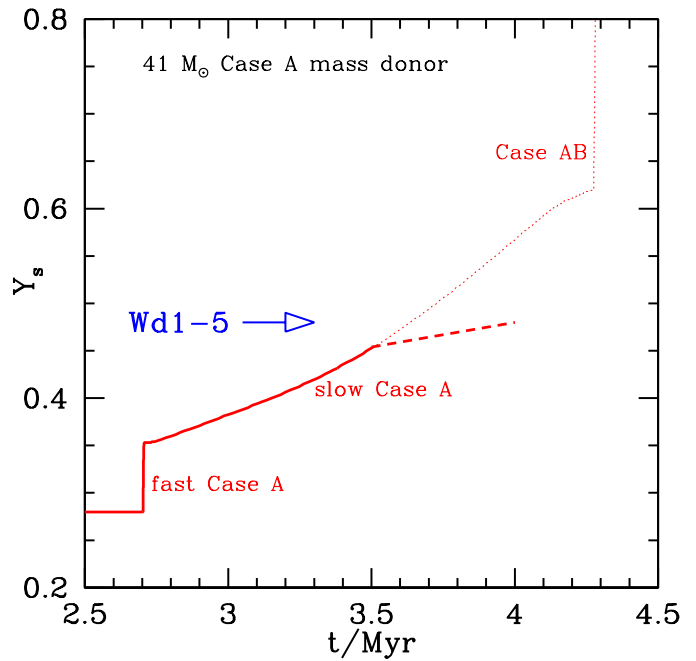}}
\caption{Evolution of the helium abundance of a $41 M_{\odot}$ primary in an initial 6~d orbital period 
massive binary under the case A
 evolutionary scenario described in Sect. 4.1, where mass loss and 
He-enrichment is driven by both stellar wind and binary induced mass transfer.
(solid red line). The dashed line represents our prediction of the evolution
of Wd1-5 in a post-common envelope phase, whereby only stellar wind mass loss is 
present; for completeness the dotted line represents the continuation 
of the specific evolutionary model of Petrovic et al. (\cite{petrovic}) if a common
envelope had not formed. Finally, a brief episode of wind-driven reverse mass transfer
during the pre-SN WC phase of the companion star accounts for the anomalous C-rich chemistry of Wd1-5.}
\end{center}
\end{figure}

At first glance, the dual requirements of binary driven mass loss and gain required to explain the abundance pattern 
of Wd1-5 appear contradictory. Nevertheless we believe they may both be accomodated in a single evolutionary scheme, which we describe here. 
We first consider the mass loss mechanism responsible for the   removal of  the majority of the  H-mantle of Wd1-5. Case B and case C mass transfer
would have quickly yielded unacceptably high surface He-abundances (e.g. $Y_s$=0.8 in 10$^4$yr for case C), which would have  subsequently  increased
very quickly, resulting in a low likelihood of catching Wd1-5 in its current state. Conversely,  while case A mass transfer also produces stars with  
He-abundances comparable to that of Wd1-5, crucially the state persists for a significant fraction of the stellar lifetime 
($\sim 10^6$ yr; Fig. 10, following Wellstein \& Langer \cite{wellstein}); we therefore conclude  that Wd1-5 
likely evolved via case A 
evolution.

The requirement for the mass-{\em gainer} to both avoid merger and explode first to unbind Wd1-5 places important constraints on the pre-SN 
system. Assuming a $\sim 41 M_{\odot}$  primary (Ritchie et al \cite{ben10}), a rather massive companion is required to avoid merger during the binary interaction, but even in a $M_{\rm init}\sim41M_{\odot}+\sim30M_{\odot}$ system - as suggested for the eclipsing BHG/WNLh+O supergiant binary  Wd1-13 (Ritchie et al. \cite{ben10}) -   we would still
expect the initial 
primary and mass donor to undergo SN first,  despite the companion accreting significant quantities of matter.
 This behaviour results from the fact that the evolution of the mass-gainer is rather sensitive to the assumed timescale
 for semiconvective mixing;  while the mass-gainer 
attempts to increase  its convective core mass in response to the accretion  of material, this rejuvenation process is 
hindered by  the chemical barrier imposed 
by the presence of (accreted) H-rich matter on top of He-rich material (see Braun \& Langer \cite{braun}). However, if the initial mass ratio of the putative 
binary system instead approached unity, a reversal of the supernova order would be expected (Wellstein et al. \cite{wellstein01}), 
while rejuvenation of the mass gainer would still not occur.  We note that such a mass ratio would 
not be expected to effect the case A evolution of the mass donor, which would still evolve into the same He-rich 
overluminous state that we currently find 
 Wd1-5 to be in.

So, we might suppose an hypothetical binary initially comprising two  $\sim 41M_{\odot}+ 35M_{\odot}$ stars - 
where the more massive component represents Wd1-5 - 
in a compact  (initial orbital period $P_{init} < \sim 8$~d) configuration in order to permit case A mass transfer.
Following the approach of Petrovic et al. (\cite{petrovic}), such 
an evolutionary pathway would lead to the $41M_{\odot}$ primary losing $\sim 20M_{\odot}$ by  $t\sim3.5$~Myr during fast case A mass transfer,
corresponding to the fast rise in $Y_s$ at that time (e.g. system \#4 of Petrovic et al. \cite{petrovic}; 
Fig. 10). The original secondary accretes a sizable fraction of this
material and becomes a $\sim 55M_{\odot}$ star. Because it is so luminous as a result of this process and as it does
not rejuvenate, it finishes core H-burning well before the donor star does (i.e. before $t=4.25$Myr). During the remaining core H-burning stage of the mass-gainer, the system is of Algol type, meaning that the mass-donor (corresponding to Wd1-5) fills 
its Roche lobe and undergoes slow case A mass transfer. During that phase, the surface helium abundance subsequently rises
 from about $Y_s=0.34$ to a value of up to $Y_s=0.47$; in excellent agreement
 to that found for Wd1-5 (Sect. 3.1).

Petrovic et al. (\cite{petrovic}) showed that the mass-gainer is significantly
spun-up by such an accretion process. In their  $56M_{\odot} + 33M_{\odot}$ system,
the mass-gainer subsequently spins down again after accretion due to its long remaining
 core hydrogen burning life time after rejuvenation. In our case, with a mass ratio 
close to 1 and without rejuvenation, the mass-gainer is not likely to  have the time to do so,
 but will instead keep a high specific angular momentum until core hydrogen exhaustion.

When the mass-gainer finishes core H-burning, the quantitative predictions of the model end.
The mass-gainer will certainly expand after core H-exhaustion and, with an orbital  period in the $\sim$10-20 
day range, it will interact with Wd1-5, likely engulfing it. While one would typically expect such interaction 
to lead to merger, we emphasise that the mass ($\sim 55 M_{\odot}$) and hence luminosity of the mass-gainer will 
be so high that its envelope will approach the Eddington limit and hence be only lightly bound. 
Therefore we suggest that both stars survive 
this phase, with the H-rich mantle of the mass-gainer ejected in a LBV/common envelope phase analagous to 
those driven by the RSG phase in lower mass systems as the orbital period of the binary decreases. This process will result in  its transition to a WR state.  Unfortunately, current simulations of binary evolution do not include such 
a phase due to the complexity of the physics.
However, in support of this assertion we emphasise that the current 
configurations of the high-mass X-ray binaries  4U1700-37 and OAO 1657-415 also require the occurrence of 
exactly such a process  in systems containing stars too 
massive to pass through a RSG phase (Clark et al. \cite{clark02}, Mason et al. \cite{mason}); 
demonstrating that such a scheme is indeed viable and has observational precedents.

 Subsequently, due to the short period of our binary - and guided by  the evolution of the analagous pre-SN binary progenitor of 4U1700-37 (Clark et al. 
\cite{clark02}) - we might expect 
 pollution of the atmosphere of Wd1-5 by the carbon-rich stellar wind of the pre-SN WC phase of the mass-gainer; this representing the second phase of (reverse) wind-driven
mass transfer required to yield the observed surface abundance pattern of Wd1-5. 
Approximately $0.05M_{\odot}$ of carbon would be required to have been accreted to 
replicate the current chemistry\footnote{Similar quantities of helium would also be transfered but would have a negligible affect on the   abundances of 
Wd1-5}; 
hydrodynamical simulations (Dessart et al. \cite{dessart}) suggest such a process is possible and it indeed appears to have
occured in 4U1700-37, despite the more powerful wind of the recipient in this system (HD~153919; 
Clark et al. \cite{clark02}).
Finally, the WC mass-gainer will  explode as a Type Ibc SN and will likely unbind the binary. The absence of rejuvenation during the case A accretion and the subsequent early exposure of the core results  in a sufficiently low pre-SN iron core mass to form a NS rather than a BH (e.g. Fryer et al. \cite{fryer}). We explore this phase of the binary evolution in more detail in Sects. 5 and 6. 

\subsection{Evolutionary pathways for massive stars in Wd1} 

Building on the evolutionary schemes delineated in Clark et al. (\cite{clark11})
 we might suppose an additional pathway for close binaries comprising two stars of comparable masses, such that for 
stars within Wd1 current evolving from the MS ($M_{\rm init}\sim35-50 M_{\odot}$):\newline

{\bf Case A binary, similar masses:} O6-7 V (primary) + O6-7 V (secondary) $\rightarrow$ case A mass transfer $\rightarrow$ 
BHG/WNLh +   O III-V  $\rightarrow$ BHG/WNLh + LBV $\rightarrow$ LBV/common envelope evolution $\rightarrow$
BHG/WNLh + WN/WC (+ wind driven mass transfer) $\rightarrow$ SN + binary disruption $\rightarrow$ BHG/WNLh + magnetar/NS 
 \newline

{\bf Case A binary, disimilar masses:} O6-7 V (primary) + OB V (secondary) $\rightarrow$ late case A/B  mass transfer
$\rightarrow$ BHG/WNLh +   OB III-V $\rightarrow$ WNo + OB III-V  $\rightarrow$ WC + OB III-V
$\rightarrow$ SN + binary disruption $\rightarrow$ NS + OB I-III \newline

{\bf Single channel:}    O6-7 V $\rightarrow$ O8-9 III $\rightarrow$ O9-B3 Ia $\rightarrow$ B5-9 Ia$^+$/YHG $\rightarrow$ 
RSG $\rightarrow$ B5-9 Ia$^+$/YHG/LBV $\rightarrow$ WN $\rightarrow$ WC(/WO?) $\rightarrow$ SN 
(leading to BH formation?)
\newline

 Comprising  two distinct populations, the BHGs within Wd~1 provide an elegant illustration of this scheme.  
The first, consisting of the late B5-9 Ia$^+$ stars Wd1-7, -33 and -42a, 
lack  observational signatures of binarity and appear to originate via single star evolution as the stars 
evolve from the main sequence and execute a red loop across the HR diagram.

The second, made up of the hybrid early BHG/WNLh  objects Wd1-5, -13 and  -44, all show clear signatures of current or historic binarity (e.g. periodic RV variability and/or hard, 
overluminous X-ray emission; Clark et al. \cite{clark08}, Ritchie et al. \cite{ben10} and  \cite{ben12}).
 Indeed, their spectral similarity (Figs. 2-4) suggest they have all 
experienced binary driven mass loss.
With an orbital  period of $\sim$9.13days, Wd1-13 is likely to 
undergo late case A/case B mass transfer and with $M_{\rm init}\sim41M_{\odot}+\sim30M_{\odot}$
will not experience the reversal of SNe order we anticipate for our putative Wd1-5 binary (Ritchie et al. \cite{ben10}).

Initial analysis of multi-epoch RV observations  of
 Wd1-44 (Ritchie et al. \cite{ben12}) suggests a period of ${\lesssim}9$d, which is potentially consistent with 
the case A mass transfer we propose for Wd1-5. Moreover, with a secondary of
apparently  earlier spectral type - and hence more massive - 
than the O supergiant companion in Wd1-13 (for which we find $M_{\rm current}\sim 35.4_{-4.6}^{+5.0}M_{\odot}$) 
it may  represent a precursor of our putative Wd1-5 binary prior to the common envelope phase
(and subsequent brief episode of reverse mass transfer that 
distinguishes Wd1-5 from the other 
two stars via the resultant C-enrichment). 

We also identify a larger population of massive compact, interacting  OB+OB 
binaries within Wd1 (e.g. Wd1-30a, -36 and -53; Clark et al. \cite{clark08}, Ritchie et al. 
\cite{ben12}) that provide a rich reservoir of progenitors from which systems such as Wd1-5, -13 \& -44   
  may be drawn. In particular we highlight 
the binary supergiant B[e] star Wd1-9, which currently  appears to be undergoing the rapid case A
 evolution we hypothesise for Wd1-5 (Fig. 10; Clark 
et al. \cite{w9}).

\section{A physical connection between Wd1-5 and CXOU J1647-45?}

Two mechanisms have been invoked to explain the runaway phenomenon; dynamical ejection from dense stellar systems;
 (Poveda et al. \cite{poveda})  and SN kicks in binary systems (Blaauw \cite{blaauw}). 
$N$-body simulations of both Wd1 and the young massive cluster R136 demonstrate that dynamical ejection of massive stars with velocities ranging up to 
$\sim300$km\,s$^{-1}$ is well underway by $\sim$3~Myr (Banerjee et al. \cite{banerjee}, Fujii et al. \cite{fujii}). Conversely,  at the age of Wd~1
  we would   expect SNe every $7-13,000$~yr ({Muno et al. \cite{muno06a}, \cite{muno06b}); 
trivially, the  presence of the magnetar confirms the recent occurrence of a SN.

Both mechanisms therefore appear viable for the ejection of Wd1-5, although a   key discriminator between dynamical and SN ejection mechanisms is 
that under the latter scenario mass transfer from the SN progenitor 
 may   result in anomalous  physical properties  of the runaway. Wd1-5 shows evidence of  C-enrichment that can only be understood via binary 
interaction, 
{\em strongly favouring the SN kick model for the formation
 of Wd1-5.} Indeed, if Wd1-5 were ejected via dynamical interaction, it must still have followed an identical 
binary evolution to that  described in Sect. 4 prior to 
this event, which would also have had to unbind the requisite evolved companion.
Therefore, while such a sequence is in principle possible, it appears unnecessarily  contrived in comparison to the SN scenario.

If the peculiar velocity imparted to  Wd1-5 was the result of a SN kick, an obvious question is whether this 
was the event that produced the magnetar J1647-45?  Given the rarity of 
magnetars, a compelling case for cluster membership rather than chance superposition may be made for  J1647-45 (Muno et al. \cite{muno06a}). 
Moreover, spectral modelling of J1647-45 shows that 
the column density towards it is consistent both with that inferred for Wd1 (determined from  optical reddening), as well as the column density towards other X-ray 
bright cluster members (Clark et al. \cite{clark08}). Following these lines of argument, we 
adopt the hypothesis that it is a cluster member for the remainder of this study.

Therefore, if  J1647-45 resides within Wd1, one would expect the putative 
pre-SN companion to also remain within the cluster; indeed if such an object were {\em not} present, the 
hypothesis that binary driven mass loss permitted the formation of a NS rather than a BH would be seriously challenged. Are the physical properties of Wd1-5 - e.g. composition, luminosity, temperature, surface gravity and
 velocity relative to J1647-45 - consistent with such an hypothesis?

The displacement between  J1645-47 and   Wd1-5 of $\sim$139'' implies a minimum
separation of $\sim$3.4~pc at a distance of 5~kpc. Assuming a characteristic 
 age of the order of 10$^4$~yr for magnetars (e.g. Kouveliotou et al. \cite{kouv}, Woods et al. \cite{woods})  leads to a  relative projected velocity 
between the two objects of $\sim$325~km\,s$^{-1}$. 
To date  direct measurements of the {\em transverse} velocities of six  magnetars have been 
made,  all of which are  encouragingly modest 
 ($\leq$350 km\,s$^{-1}$)\footnote{$v=212\pm{35}(d/3.5{\rm kpc})$km\,s$^{-1}$  for
 XTE J1810-197 (Helfand et al. \cite{helfand}), $v=280^{+130}_{-120}$km\,s$^{-1}$ for PSR J1550-5418 (Deller et al. \cite{deller}), $v=350\pm{100}(d/9{\rm 
kpc})$km\,s$^{-1}$ 
for SGR 1806-20 and $v=130\pm{30}(d/12.5{\rm kpc})$km\,s$^{-1}$ for SGR 1900+14 (both Tendulkar et al. \cite{tendulkar})
and $v=157\pm17$km\,s$^{-1}$  for AXP 1E 2259+586 and $v=102\pm26$km\,s$^{-1}$ for AXP 4U 0142+61 (both Tendulkar et al. 
\cite{tendulkar13}).}
 and hence consistent with the above estimate
for the J1645-47/Wd1-5 system. Moreover, the association of three anomalous X-ray pulsars with supernovae remnants 
further supports the adoption of  rather low magnetar kick velocities (despite theoretical predictions to the contrary; Mereghetti \cite{mereghetti}).
 For comparison, a mean velocity of  $\sim400$km\,s$^{-1}$ has been reported for young ($<3$Myr) 
pulsars, with a maximum velocity of $\sim1600$km\,s$^{-1}$ (Hobbs et al. \cite{hobbs}). Given these findings, the separation of both Wd1-5 and J1645-47 
appears
consistent with a common origin.

Moreover, as  highlighted in Sect. 3.2, the  anomalous C-abundance of Wd1-5 
(C/C$_{\odot} \sim 1.4$)
is also present in the B1 Ia$^+$ and O6.5 Iaf$^+$ hypergiant mass donors in the X-ray binaries GX301-2 
and 4U1700-37. As with our putative  Wd1-5 + J1647-45 binary, the current physical properties of both 
binaries imply pre-SN mass transfer onto the 
current mass donor (Wellstein \& Langer \cite{wellstein}, 
Clark et al. \cite{clark02}, Kaper et al. \cite{kaper}).
Additionally,  the lifetime of the C-abundance anomaly is expected to be rather short. Critically, acting on the 
{\em thermal} timescale (e.g. Wellstein et al. \cite{wellstein01}, Petrovic et al. \cite{petrovic}), thermohaline, 
rather than rotational, mixing is expected to act to rapidly dilute the carbon overabundance resulting from the pre-SN mass transfer. So one would expect
 that dilution would already be well advanced after only $10^4$yr; a timescale directly comparable to the lifetime 
inferred for  magnetars. Indeed, the similarity in the timescale for carbon dilution to the lifetime of a magnetar implies 
that the cessation of mass transfer to Wd1-5 and the event that formed  J1647-45 must have occurred
  quasi-simultaneously. 

We can also advance three additional arguments to bolster this association. No 
other magnetar or young cooling neutron  star candidate that could have formed the requisite companion to Wd1-5 has been identified in either the original dataset  or subsequent multi-epoch
X-ray observations (Muno et al. \cite{muno06a}, Clark et al. \cite{clark08}, Woods et al. \cite{woods11}). Conversely, despite extensive optical and near-IR surveys (Negueruela et al. \cite{iggy10}, Crowther et al. \cite{crowther}),
 no other plausible pre-SN binary companion to J1647-45 has been identified within the massive stellar population of Wd1.
Moreover, given the preceding estimate of the mean interval between consecutive SNe within Wd1 at this
 epoch, one would expect the observed runaway velocity of Wd1-5 to carry it beyond the cluster confines before 
the next such event - i.e. if the velocity of Wd1-5 is representative of that  imparted to runaway stars at SN,  
on average  one would expect  only one such object  to be present within Wd1 at this time - and hence on 
statistical grounds we would expect both magnetar and Wd1-5 to be physically associated with one another.

 So in summary, all the available evidence points to the anomalous RV of  Wd1-5 as being the result of a SNe kick. In particular, the carbon abundance points to a 
recent episode of mass transfer from a close companion, although {\em currently} there is no observational evidence of such an object.
  Moreover, the timescale for dilution of the abundance  anomalies is comparable to  the duration of the magnetar phase, while 
 the angular separation of both Wd1-5 and J1647-45 is also consistent with known magnetar and pulsar kick velocities. Therefore, 
while the absence of transverse velocity measurements prevents a definitive association,  we can identify strong lines of argument to posit a physical association between Wd1-5 and J1647-45 in a pre-SN binary system. 

\section{Implications for magnetar formation}

If  the hypothesis  that  Wd1-5 and the magnetar J1647-45 comprised a pre-SN binary system is correct, what are the physical implications?
No consensus yet exists on the formation mechanism for  magnetars. Gaensler et al.  (\cite{gaensler})
suggested their progenitors were limited to particularly  massive stars ($\gtrsim 40M_{\odot}$); however,
more recent observations  suggest  that they instead span a wide range of masses ($\sim17-50 M_{\odot}$; 
Sect. 1), implying that additional physical factors must drive this process.

Duncan \& Thompson (\cite{duncan}) and Thompson \& Duncan (\cite{thompson}) 
argued for their formation  via rapidly rotating ($P \sim 1$ms) proto-NSs, when a 
large-scale convective dynamo may generate an extreme magnetic field in the first few seconds after 
birth. However, if  magnetic torques are successful at removing angular momentum from 
the core via coupling to the extended atmosphere present in a pre-SN RSG phase, then 
the core will not be rotating  rapidly enough at SN for this mechanism to operate
(e.g. Heger et al \cite{heger}). Moreover such a scenario would predict both highly energetic SNe and 
 high spatial velocities for the resultant magnetars (e.g. Duncan \& Thompson \cite{duncan}) which appear to be in 
conflict with  observations (e.g. Mereghetti \cite{mereghetti}, Vink \& Kuiper 
\cite{vink}). An alternative suggestion is the `fossil field' hypothesis, whereby a pre-existing  magenetic 
field acquired  at the birth of the progenitor  is amplified during stellar collapse. 
Spruit (\cite{spruit}) argued against this mechanism 
 given the lack of sufficient numbers of highly magnetised stars to explain the expected formation rate of
 magnetars and, once again, the fact that  core-envelope coupling will result in spin down of the core.

How does  J1647-45 inform the debate? The Wd1 RSGs cohort
implies that {\em if} the progenitor of J1647-45 had been a single star, it would have passed through such a phase and hence 
have  been subject to spin-down via core-envelope coupling.
While we cannot exclude an unusually strong (fossil) magnetic field in 
the progenitor of  J1647-45, we find no evidence of a corresponding  population of highly magnetic 
massive stars within Wd1 at this time\footnote{By analogy to the optical properties of Of?p stars, Clark et al. 
(\cite{clark10}) cited Wd1-24 as a possible highly magnetic star,  based on the  variable C\,{\sc iii}+Pa 16 
$\sim8500${\AA} blend. However, further observations have revealed that this behaviour 
is instead due to a variable  contribution from a late-O  binary companion (Ritchie et al. \cite{ben12}).
Known magnetic OB stars are 
 expected to demonstrate hard, overluminous X-ray emission (e.g. Clark 
et al. \cite{clark09} and refs. therein); while several overluminous OB  stars 
are present within Wd~1 (Clark et al. \cite{clark08}), they all demonstrate rather soft X-ray spectra; 
the sole exception being W30a, a known (colliding wind) binary.}.

Binarity has been invoked under both scenarios as an additional ingredient in order
to avoid core-envelope coupling, by   removing the outer layers of the SN progenitor  and thus preventing a RSG  
phase, so  that sufficient  angular momentum is retained  in the core to form a magnetar\footnote{Although simulations by Yoon et al. (\cite{yoon})  suggests that such a mechanism is ineffective for stars 
below $25 M_{\odot}$.}.}  Moreover, the high mass 
implied for the progenitors of both J1647-45 and SGR1806-20 (Sect. 1) suggests that binary driven mass loss resulting 
in the early onset of WR mass loss rates and hence the production of a low-mass pre-SN core was likewise essential 
to the formation of a magnetar rather than a BH (cf. Fryer et al. \cite{fryer}).

An additional  point of interest is whether the pre-SN evolution of a  massive compact binary system - 
such as we infer for J1647-45 - favours the production of a seed magnetic field, which is subsequently amplified during or 
shortly after core collapse to yield a magnetar (e.g. Spruit \cite{spruit}). Following Langer (\cite{ARAA}), 
both mass transfer and stellar merger in compact binaries may lead to dramatic spin-up of the mass-gainer/merger 
remnant, potentially favouring the formation of a magnetic field via dynamo action.
In support of such an hypothesis we  highlight the detection of a magnetic field in the rapidly rotating secondary 
in Plaskett's star (Grunhut et al. \cite{grunhut}); a higher mass analogue of our putative Wd1-5+J1647-45 binary, 
which has also undergone case A mass transfer (e.g. Linder et al. \cite{linder}).

Alternatively, Tout et al.  (\cite{tout})   suggested that  high magnetic field white dwarfs may form  via strong binary interactions between a main sequence and red giant (and hence white-dwarf progenitor) in a 
common envelope phase.  In this scenario as the orbital period of the 2 components decreases, differential rotation 
 within the convective common envelope generates  the magnetic field via dynamo action; one might speculate that a comparable 
 process also occurs in high mass analogues, such as  the pre-SN LBV/common envelope phase we propose here (Sect.  4.1).

A binary mediated formation scenario would suppose that if merger is 
avoided\footnote{Under such a scenario we might suppose the lack of a luminous stellar 
source associated with the wind blown bubble hosting the magentar 1E 1048.1-5937
(Gaensler et al. \cite{gaensler}) would be explicable if the binary merged prior to SN.} one could anticipate identifying
 the pre-SN magnetar companion. 
The proper motion study of Tendulkar et al. (\cite{tendulkar}; footnote 6) places the birthsite of SGR1806-20 within the 
confines of the eponymous host 
cluster, with the positional uncertainty  elipse encompassing three stars. Of these, object D has been classified as an OB supergiant, and 
hence potentially the massive pre-SN companion we predict; classification spectroscopy of the remaining stars in combination with 
an RV survey of the full cluster population would be of considerable interest to determine if any of the three are indeed
 overluminous  chemically peculiar runaway analogues to Wd1-5. Alternatively, if disruption is avoided,
binaries containing magnetars should also exist. In this  regard the  suggestion of Reig et al. (\cite{reig})
that the neutron star within the  X-ray binary 4U2206+54 is a magnetar is of considerable interest, moreso 
given that the combination of short ($P_{orb}\sim10$~day) period and He-rich nature of the primary is suggestive of
 pre-SN binary interaction in the system.

 Potential observational biases in the detection of
 quiescent magnetars hamper a {\em direct}  determination of their birth rate,
although several authors have suggested it may be comparable to that 
of radio pulsars  (e.g. Muno et al. \cite{muno08}, Woods \cite{woods08});
confirmation of a binary channel for  magnetar formation 
would allow us to address this issue.
Intriguingly, recent observations suggest a high ($\geq 40$\%; e.g.  Sana et al. \cite{sana12},
 Kiminki \& Kobulnicky \cite{kiminki}, Chini et al. \cite{chini}, Ritchie et al. \cite{ben12})
 binary fraction amongst OB stars, an orbital period  distribution  favouring short-period systems 
with respect to the  classical \"{O}pik's Law (a flat distribution of orbital 
separations in logarithmic space) and a mass ratio favouring more massive companions (i.e. inconsistent with random 
selection from a Kroupa type initial mass function); all factors potentially favouring the production of magnetars under the above scenario.
Indeed, the presence of a number of compact OB+OB binaries 
and a massive binary fraction of $>70$\% 
amongst the WR population (Clark et al. \cite{clark08})
potentially provides a rich progenitor reservoir within Wd1, with 
Wd1-44 being the most compelling example currently identified (Sect. 4.2).

\section{Concluding remarks}
 The presence of a magnetar with a progenitor mass $\gtrsim 40M_{\odot}$ within Wd~1 requires a mechanism by which significant
 mass loss can occur prior to SN, with binary interaction a leading candidate. Therefore, the identification of a pre-SN companion to J1647-45  provides a critical test of the theory.  As part of our RV survey of the cluster we identified the single star Wd1-5 as a runaway - and hence a potential candidate.

To date no other massive star - either within  Wd1 or part of the wider Galactic population - completely reproduces 
the spectral morphology and/or combination of physical properties of Wd1-5. 
Quantitative analysis reveals physical properties inconsistent with the evolution of a single star (Sect. 3). In particular the anomalously high C-abundance of Wd1-5 in comparison to the predicted CNO abundances is inexplicable under such a scenario and  has previously  only been observed in  the mass donors of the X-ray binaries GX301-2 and 4U1700-37. In both  cases this is  thought to result  from a brief episode  of wind driven mass transfer from a C-rich WC Wolf-Rayet binary companion which, post-SN, formed the relativistic companion.

Motivated by these findings, we used the combination of spectroscopic mass, luminosity and chemical 
abundances of Wd1-5 to infer a pre-SN 
evolutionary history for the putative binary (Sect. 4). Significant case A mass transfer from Wd1-5 leads to spin-up of the
companion, which consequently evolves more rapidly than the mass donor, resulting in a subsequent LBV-driven common-envelope phase which 
strips its H-rich mantle. The initially less massive companion then enters the WR phase, triggering an enrichment of the atmosphere of 
Wd1-5 by the stellar wind of the by then WC  star, which then explodes as a type Ibc SN, unbinding the binary. In support of this hypothesis we highlight that the observational properties of the remaining early-BHG/WNLhs within Wd1 - Wd1-13 and -44 - reveal them both to be short-period interacting binaries, while a large reservoir of progenitor binaries has also been identified (Ritchie et al. \cite{ben12}).

Given this, a natural explanation for the runaway nature for Wd1-5 is that it was ejected via a SN kick and we present a number of lines
of argument to support a physical association with the magnetar J1647-45 (noting that the conclusions above are {\em not} dependant on such a connection). Under this hypothesis, binarity  plays a critical role in the formation of the magnetar by  
(i)  preventing the spin-down which happens in single stars via core-envelope 
coupling because the envelope is removed and (ii) enabling the 
formation of a low-mass pre-SN core via the prolonged action of WR-phase winds. Moreover, we
may speculate that the binary interaction results in the generation of a seed magnetic field in the 
magnetar progenitor via dynamo action, either during the spin up of the mass gainer or in a subsequent LBV/common envelope 
phase. 

If correct, while the BHG/WNLh stars Wd1-5, -13 and -44 might be expected to form a NS rather than a BH as a result of their binary driven mass loss, they have not been spun up by mass transfer and therefore do not replicate the properties of our putative magnetar progenitor. While the secondary  in Wd-13 has been spun up via mass transfer, it remains insufficiently massive for the SN order to be reversed, as we propose for Wd1-5, and consequently will follow a different evolutionary path whereby
magnetic core-envelope coupling is not avoided. However,
the secondary in Wd1-44 appears more massive and we might suppose this system will 
follow a similar pathway to Wd1-5, and hence potentially yield a magnetar. 

As well as  permitting a determination of the magnetar formation rate via the identification of the  
binary progenitor population, the presence of carbon pollution in the atmosphere of Wd1-5 supposes an H-depleted  
WC Wolf-Rayet as the immediate magnetar  progenitor, which would  have exploded as a type Ibc SN. 
This would be the first association of the birth of a magnetar with such an event and would also support 
the assertion that massive close binary evolution is a promising channel for the production of a subset of  type Ibc SNe. 
Indeed current observational studies suggest that binary stripping is important in the production of the majority of  
type Ibc SNe, albeit  from a population of  lower mass 
($\leq 20-25M_{\odot}$) progenitors than we assume here (Smith et al. \cite{smith}, Eldridge et al. \cite{eldridge}, Kuncarayakti et al. \cite{kun}).

Expanding upon this and a number of authors have suggested that 
magnetars may power superluminous type II and Ibc SNe  (Thompson et al. \cite{thompson04}, Woosley \cite{woosley10}, Kasen \& Bildsten \cite{kasen}, Gal-Yam \cite{galyam} and Quimby et al. \cite{quimby}); indeed the 7.29M$_{\odot}$ progenitor 
 model of Woosley (\cite{woosley10}) is directly motivated by the presence of J1647-45 within Wd1. Moreover, Inserra et al. (\cite{inserra}) studied the  late-time
lightcurves of five superluminous type Ic SNe, finding that the data are indeed consistent with these events
being powered by the rapid spin-down of newly born magnetars (see also McCrum et al. \cite{mccrum} and Nicholl et al. \cite{nicholl})
Additionally, magnetars have also been proposed as the central engines  of some gamma-ray bursts
(GRBs; e.g. Usov \cite{usov}, Duncan \&  Thompson \cite{duncan} and Metzger et al. \cite{metzger} 
and refs. therein). Given the latter suggestion it is therefore intriguing that  long  duration GRBs have been associated with  type Ibc 
SNe   (e.g. Della Valle \cite{dv}).

If such 
an hypothesis is viable, one might ask why such superluminous events  are not more common in the local Universe
 if the magnetar formation rate is indeed a substantial percentage of  that of  neutron stars (e.g. Muno et al. 
\cite{muno08})? One plausible  explanation may be that the superluminous SNe occur in low metallicity environments 
where correspondingly weak stellar winds minimise pre-SNe angular momentum losses, leading to systematically more
 rapidly rotating magnetars and  hence a greater deposition of energy in the SNe  
in comparison  to the higher metallicity local  environment. In any event, given the apparent ubiquity 
of massive compact binaries, additional theoretical and observational
investigations of  the potential link between binary mediated formation 
channels for magnetars, (superluminous) type Ibc SNe and GRBs would clearly be of considerable interest.

 \begin{acknowledgements}
This research is partially supported by the Spanish Ministerio de Ciencia e Innovaci\'on (MICINN) under 
grants AYA2010-21697-C05-01, AYA2012-39364-C02-02 and FIS2012-39162-C06-01.

\end{acknowledgements}

{}

\appendix
\section{Summary of stellar comparison data to Wd1-5}

\longtab{1}{
\begin{longtable}{llccccccc}
\caption{Comparison of basic stellar parameters of Wd1-5 to those of related galactic stars.}\\
\hline\hline
 Name & Spec.& log(\Lstar) & \Rstar     & \Teff &  \logg    & log(\Mdot) & \vinf & Reference\\
   &Type  & ($L_{\odot})$       &  (\Rsun)     &  (kK) &    & ($M_{\odot}$yr$^{-1}$) & (\kms) &  \\
\hline 
\endfirsthead
\caption{continued.}\\
\hline\hline
Name & Spec.& log(\Lstar) & \Rstar     & \Teff &  \logg    & log(\Mdot) & \vinf & Reference\\
   &Type  & ($L_{\odot}$)       &  (\Rsun)     &  (kK) &    & ($M_{\odot}$yr$^{-1}$) & (\kms) &  \\
\hline
\endhead
\hline
\endfoot
{\bf Wd1-5}  &B0.5 Ia$^+$ &  $5.38^{+0.12}_{-0.12}$ &  $34.0^{+5.0}_{-4.4}$ &  $21.05^{+1.5}_{-1.2}$ & 
 $2.33^{+0.17}_{-0.10}$ &  $-5.36^{+0.30}_{-0.20}$ &  $430^{+20}_{-40}$
&  This work \\[2.5mm]
HD 30614 &O9.5 Ia     & 5.63 & 26.0  & $29.0\pm1.0$ & $3.0\pm0.15$ & $-5.30^{+0.11}_{-0.15}$ & 1560 & 1\\
HD 168183&O9.5 Ib      &$5.42\pm0.22$ & $19.0^{+4.3}_{-3.5}$     & $30.0\pm1.0$     & $3.3\pm0.1$    & $-6.82^{+0.24}_{-0.24}$     & $1700\pm510$    & 2 \\
HD 37128 &B0 Ia     & 5.44     &  24.0    & $27.0\pm1.0$     & $2.9\pm0.15$    & $-5.60^{+0.11}_{-0.15}$     & 1910    & 1 \\
HD 89767   & B0 Ia  & $5.35\pm0.22$    & $30.0^{+6.7}_{-5.5}$     & $23.0\pm1.0$     & $2.55\pm0.1$    & $-6.07^{+0.24}_{-0.24}$    & $1600\pm480$    & 2 \\
HD 91969 &B0 Ia     & 5.52     &  25.3    & $27.5\pm1.0$     & $2.75\pm0.14$    & $-6.00^{+0.11}_{-0.15}$     & 1470    & 1 \\
HD 94909 &B0 Ia     & 5.49     &  25.5    & $27.0\pm1.0$     & $2.9\pm0.14$   &  $-5.70^{+0.11}_{-0.15}$   & 1050    & 1 \\
HD 122879&B0 Ia     & 5.52     &  24.4    & $28.0\pm1.0$     & $2.95\pm0.14$    & $-5.52^{+0.11}_{-0.15}$     & 1620    & 1 \\
HD 192660&B0 Ib     & $5.74\pm0.13$     &  $23.4\pm1.0$   &  $30.0\pm1.0$ & 3.25 & $-5.30^{+0.00}_{-0.40}$     &1850    & 3 \\
HD 204172&B0.2 Ia     & $5.48\pm0.27$     &  $22.4\pm3.2$   &  $28.5\pm1.0$ & 3.13 &  $-6.24^{+0.34}_{-0.40}$ & 1685   & 3 \\
HD 38771 &B0.5 Ia     & 5.35     &  22.2    & $26.5\pm1.0$     & $2.9\pm0.14$    &  $-6.05^{+0.11}_{-0.15}$   & 1525    & 1 \\
HD 115842&B0.5 Ia     & 5.65     &  34.2    & $25.5\pm1.0$     & $2.85\pm0.14$    & $-5.70^{+0.11}_{-0.15}$    & 1180    & 1 \\
HD 152234& B0.5 Ia     &  5.87     &  42.4    & $26.0\pm1.0$     & $2.85\pm0.14$    & $-5.57^{+0.11}_{-0.15}$     & 1450    & 1 \\
HD 185859&B0.5 Ia  & $5.54\pm0.14$ &  $29.1\pm1.3$   &  $26.0\pm1.0$    & 3.13    & $-6.30^{+0.08}_{-0.10}$     &1830    & 3 \\
HD 64760&B0.5 Ib & $5.48\pm0.26$    &  $23.3\pm2.2$   &  $28.0\pm2.0$    & 3.38    & $-5.96^{+0.28}_{-1.04}$     & 1600   & 3 \\
HD 93619 &B0.5 Ib     &$5.30\pm0.22$     &   $22.0^{+4.9}_{-4.1}$    & $26.0\pm1.0$ &    $2.9\pm0.1$     & $-6.12^{+0.24}_{-0.24}$      & $1470\pm441$   & 2 \\ 
HD 213087&B0.5 Ib  & $5.69\pm0.11$     &  $32.0\pm0.01$   &  $27.0\pm1.0$    & 3.13    & $-6.15^{+0.19}_{-0.0}$ &1520    & 3 \\HD 2905  &BC0.7 Ia     &  5.52    &   41.4   &  $21.5\pm1.0$    & $2.6\pm0.14$    &  $-5.70^{+0.11}_{-0.15}$   & 1105    & 1 \\
HD 91943 &B0.7 Ia     & 5.35     &  26.3    & $24.5\pm1.0$     &$2.8\pm0.14$     & $-6.12^{+0.11}_{-0.15}$     & 1470    & 1 \\
HD 152235&B0.7 Ia     &  5.76    & 47.1     &$23.0\pm1.0$      & $2.65\pm0.14$    &  $-5.90^{+0.11}_{-0.15}$    &850     & 1 \\
HD 154090&B0.7 Ia     & 5.48     & 36.0     & $22.5\pm1.0$     &$2.65\pm0.14$     & $-6.02^{+0.11}_{-0.15}$     & 915    & 1 \\
HD 96880 & B1 Ia     & $5.42\pm0.11$    & $43.0^{+9.7}_{-7.9}$    &$20.0\pm1.0$ & $2.4\pm0.1$    & $-6.40^{+0.24}_{-0.24}$    & $1200\pm360$ & 2 \\
HD 115363 & B1 Ia    & $5.42\pm0.11$    & $43.0^{+9.7}_{-7.9}$     & $20.0\pm1.0$ & $2.4\pm0.1$   & $-5.92^{+0.24}_{-0.24}$ & $1200\pm360$ & 2 \\
HD 148688& B1 Ia     & 5.45      &36.7      &$22.0\pm1.0$      &$2.60\pm0.14$     & $-5.76^{+0.11}_{-0.15}$     & 725    & 1 \\
HD 170938  & B1 Ia & $5.42\pm0.11$ & $43.0^{+9.7}_{-7.9}$ & $20.0\pm1.0$ & $2.4\pm0.1$ & $-6.15^{+0.24}_{-0.24}$ & $1200\pm360$ & 2 \\
HD 13854 &B1 Iab     & 5.43     & 37.4     & $21.5\pm1.0$     &$2.55\pm0.14$     & $-6.07^{+0.11}_{-0.15}$     & 920    & 1 \\
HD 91316 &B1 Iab     & 5.47     & 37.4     & $22.0\pm1.0$     &$2.55\pm0.14$     & $-6.46^{+0.11}_{-0.15}$     & 1110    & 1 \\
HD 109867 & B1 Iab   & $5.56\pm0.22$ & $38.0^{+8.5}_{-7.0}$ & $23.0\pm1.0$  & $2.6\pm0.1$ & $-6.30^{+0.24}_{-0.24}$ & $1400\pm420$ & 2 \\
HD 190066&B1 Iab  & $5.54\pm0.20$  &  $41.4\pm1.9$   &  $21.0\pm1.0$    & 2.88    & $-6.15^{+0.05}_{-0.07}$     & 1275   & 3 \\
HD 47240 & B1 Ib       & $4.93\pm0.22$ & $27.0^{+6.1}_{-5.0}$ & $19.0\pm1.0$  & $2.4\pm0.1$ & $-6.77^{+0.24}_{-0.24}$ & $1000\pm300$ & 2 \\
HD 154043 &B1 Ib & $4.98\pm0.22$      & $26.0^{+5.8}_{-5.3}$ & $20.0\pm1.0$  &$2.5\pm0.1$ & $-6.70^{+0.24}_{-0.24}$ &$1300\pm390$ &2 \\
HD 54764 & B1 Ib/II   & $4.90\pm0.22$ & $26.0^{+5.8}_{-5.3}$ & $19.0\pm1.0$  & $2.45\pm0.10$ & $-7.52^{+0.24}_{-0.24}$ & $900\pm270$ & 2 \\
HD 14956 &B1.5 Ia     & 5.65     &50.6      & $21.0\pm1.0$     &$2.5\pm0.14$     &  $-6.00^{+0.11}_{-0.15}$    & 500    & 1 \\
HD 106343 & B1.5 Ia   & $5.40\pm0.22$ & $42.0^{+9.4}_{-7.7}$ & $20.0\pm1.0$  & $2.50\pm0.1$ & $-6.28^{+0.24}_{-0.24}$ & $800\pm240$ & 2 \\
HD 193183 &B1.5 Ib & $5.00\pm0.26$  &  $30.8\pm2.8$   &  $18.5\pm1.0$    & 2.63    &  $-6.64^{+0.40}_{-0.00}$    & 565   & 3 \\
HD 111990 & B1/2 Ib  &$4.91\pm0.22$ & $25.0^{+5.6}_{-4.5}$ & $19.5\pm1.0$  & $2.55\pm0.10$  &$-6.85^{+0.24}_{-0.24}$ & $750\pm225$ & 2 \\
HD 14143 &B2 Ia      & 5.42     &52.9      & $18.0\pm1.0$     &$2.25\pm0.14$     & $-5.98^{+0.11}_{-0.15}$     &645     & 1 \\
HD 14818 &B2 Ia     & 5.35     &46.1      & $18.5\pm1.0$     &$2.4\pm0.14$     & $-6.26^{+0.11}_{-0.15}$     & 565    & 1 \\
HD 41117 &B2 Ia     & 5.65     &61.9      & $19.0\pm1.0$     &$2.35\pm0.14$     & $-6.05^{+0.11}_{-0.15}$     & 510    & 1 \\
HD 194279&B2 Ia     & 5.37     &44.7      & $19.0\pm1.0$     &$2.3\pm0.14$     & $-5.98^{+0.11}_{-0.15}$     & 550    & 1 \\
HD 206165&B2 Ib  & $5.18\pm0.26$  &  $39.8\pm5.5$   &  $18.0\pm0.5$    & 2.50    & $-6.30^{+0.00}_{-0.22}$     & 640   & 3 \\
HD 141318 &B2 II & $4.56\pm0.11$ & 16.0 & $20.0\pm1.0$ & $2.90\pm0.1$ & $-7.52^{+0.24}_{-0.24}$ & $900\pm270$ & 2 \\    
HD 92964 &  B2.5 Iae & $5.33\pm0.11$ & 48.0 & $18.0\pm1.0$  & $2.1\pm0.1$ & $-6.55^{+0.24}_{-0.24}$& $520\pm156$ & 2\\
HD 198478&B2.5 Ia     & 5.03      &40.0      &$16.5\pm1.0$      &$2.15\pm0.14$     & $-6.64^{+0.11}_{-0.15}$     &470     & 1 \\
HD 42087&B2.5 Ib & $5.11\pm0.24$     &  $36.6\pm1.7$   &  $18.0\pm1.0$    & 2.50    &  $-6.70^{+0.00}_{-0.40}$    &650    & 3 \\HD 14134 &B3 Ia     & 5.28      &56.7      &$16.0\pm1.0$      & $2.05\pm0.14$    & $-6.28^{+0.11}_{-0.15}$    &465     & 1 \\
HD 53138 &B3 Ia     & 5.34      &65.0      & $15.5\pm1.0$     & $2.05\pm0.14$    & $-6.44^{+0.11}_{-0.15}$    &865    & 1 \\
HD 58350&B5 Ia     & $5.18\pm0.17$     &  $57.3\pm2.6$   &  $15.0\pm0.5$    & 2.13    &  $-6.15^{+0.15}_{-0.00}$  & 320   & 3 \\
HD 102997 &B5 Ia & $5.25\pm0.22$ & $55.0^{+12.4}_{-10.1}$ & $16.0\pm1.0$ & $2.00\pm0.1$ & $-6.64^{+0.24}_{-0.24}$ & $325\pm98$ & 2 \\
HD 108659 &B5 Ib & $4.60\pm0.22$ & $26.0^{+5.8}_{-5.3}$ & $16.0\pm1.0$  & $2.30\pm0.1$ & $-7.22^{+0.24}_{-0.24}$ & $470\pm141$ & 2 \\
HD 164353 &B5 Ib     & $4.30\pm1.30$     &  $19.6\pm8.1$   &  $15.5\pm1.0$    & 2.75    & $-7.22^{+0.00}_{-0.30}$  &  450  & 3 \\HD 191243&B5 Ib     & $5.30\pm0.37$     &  $70.8\pm3.3$   &  $14.5\pm1.0$    & 2.75    & $-6.08^{+0.00}_{-0.56}$  & 550   & 3 \\
HD 80558 & B6 Iab & $4.78\pm0.22$ & $45.0^{10.1}_{-8.3}$ & $13.5\pm1.0$ & $1.75\pm0.1$ & $-7.30^{+0.24}_{-0.24}$ & $250\pm75$ & 2 \\
HD 91024 & B7 Iab & $4.74\pm0.22$ & $50.0^{-11.4}_{-9.1}$ & $12.5\pm1.0$ & $1.95\pm0.1$ & $-7.00^{+0.24}_{-0.24}$& $225\pm68$ & 2 \\
HD 199478&B8 Iae  & 5.08 & 68.0 & $13.0\pm1.0$ & 1.70 & -6.70...-6.15 & 230 & 4\\
HD 94367 & B9 Ia & $4.97\pm0.22$ & $77.0^{+17.3}_{-14.1}$ & $11.5\pm1.0$ & $1.55\pm0.1$ & $-6.62^{+0.24}_{-0.24}$ & $175\pm53$ &2 \\
HD 202850& B9Iab & $4.59\pm0.22$ & 54.0 & $11.0\pm1.0$ & 1.85 & $-7.22^{+0.26}_{-0.18}$ & 240 & 4 \\
HD 212593&B9 Iab & $4.79\pm0.22$  & 59.0 & $11.8\pm1.0$ & 2.18 & $-7.05^{+0.25}_{-0.17}$ & 350 & 4 \\[2.5mm]

BP Cru     & B1 Ia$^+$  & 5.67&70.0 & $18.1^{+0.5}_{-0.5}$& 2.38& -5.00 & 305 & 5\\
$\zeta^1$ Sco& B1.5 Ia$^+$  & 5.93&103.0 & $17.2\pm0.5$ &1.97 & -5.2 & $390\pm50$  & 6\\
HD~190603 & B1.5 Ia$^+$    & 5.58&63.0 & $18.0\pm0.5$& 2.10& -5.66 &$485\pm50$ & 6\\
Cyg \#12   & B3-4 Ia$^+$  & 6.28&246.0 & $13.7^{+0.8}_{-0.5}$ & $1.70^{+0.08}_{-0.15}$ &-4.82 & $400^{+600}_{-100}$ & 6\\[2.5mm]
AG Car & LBV & $6.17^{+0.04}_{-0.05}$ & 58.5 & $22.8\pm0.5$ & - & $-4.22^{+0.11}_{-0.16}$ & $300\pm30$ & 7 \\
       &     & $6.00^{+0.04}_{-0.05}$ & 115.2 & $14.3\pm0.5$ & - & $-3.92^{+0.11}_{-0.16}$ & $150\pm30$  & 7 \\
P Cygni& LBV & 5.85                   & 76.0  & 19.2         & - & -4.49                   & 185         & 8 \\
HR Car & LBV & 5.70                  & 70.0  & 17.9         & - & -4.85                   & 120         & 9 \\
Pistol star & LBV & 6.20 & 306 & $11.8\pm1.5$ & - & -4.12 & 105 & 10\\
FMM 362 & LBV & 6.25 & 350 & $11.3\pm1.5$ & - & -4.37 & 170 & 10 \\
AFGL2298&LBV & 6.30                  & 385   & $11.0\pm0.5$ &  - & -4.28                   & 125         & 11 \\[2.5mm]

NS4 & WNLh & 5.58 & 22.6& 28.4&  - &  -4.35 & 700 & 12\\
HD~313846 & WNLh& 5.82& 33.1&  27.7 &  - & -4.53 & 1170 & 12\\
GC AF & WNLh  & $5.3\pm0.2$ & $28.1\pm0.6$&21.0 & - & $-4.25\pm0.2$ & $700\pm100$ & 13\\
GC IRS16C  & WNLh & $5.9\pm0.2$& 63.9& $19.5\pm6.0$& - &$-4.65\pm0.2$ & $650\pm100$ & 13\\
GC IRS34W & WNLh& $5.5\pm0.2$&35.9 &$19.5\pm6.0$ &  - & $-4.88\pm0.2$ & $650\pm100$ & 13\\
GC IRS33E& WNLh & $5.75\pm0.2$& 63.9&$18.0\pm6.0$ & - & $-4.80\pm0.2$& $450\pm100$ & 13\\
GC IRS19NW & WNLh & $5.9\pm0.2$& 59.1&$17.5\pm6.0$ & - & $-4.95\pm0.2$ &$600\pm100$ &  13\\

\hline
\end{longtable}
{Stars are presented from early to late spectral types for BSGs and BHGs  and from hottest to coolest for the LBVs and WNLh stars. The lack 
of suitable  absorption profiles prevents a determination of the surface gravity of the LBVs and WNLh stars.
Throughout the table we employ the clumping corrected mass loss rate ({\Mdot}/$\sqrt{f}$) to
 allow direct comparison between individual stars; hence the difference between this table and Table 1 for Wd1-5. Unfortunately errors are not presented for all physical parameters of 
all individual stars by the studies used in the construction of this table. Values for both high and low 
temperature states of AG Car are provided. References are $^1$Crowther et al. (\cite{pacBSG}),
 $^2$Lefever et al. (\cite{lefever}),
 $^3$Searle et al. (\cite{searle}),
$^4$Markova \& Puls (\cite{markova}), 
$^5$Kaper et al. (\cite{kaper}), 
$^6$Clark et al. (\cite{clark12}),
$^7$Groh et al. (\cite{groh}),
$^{8}$Najarro (\cite{paco01})
$^9$Groh et al. (\cite{grohb}),
$^{10}$Najarro et al. (\cite{paco09}),
$^{11}$Clark et al. (\cite{clark09b}),
$^{12}$Bohannan \& Crowther (\cite{bohannan}) and
$^{13}$Martins et al. (\cite{martinsGC}).}
}


\begin{thebibliography}{}

\bibitem[1989]{anders}
Anders, E. \& Grevesse, N. 1989, GeCoA, 53, 197

\bibitem[1998]{appenzeller}
Appenzeller, I., Fricke, K., F\"urtig, W., et al. 1998, The Messenger, 94, 1

\bibitem[2006]{asplund}
Asplund, M., Grevesse, N. \& Jacques Sauval, A. 2006, NuPhA, 777, 1



\bibitem[2012]{banerjee}
Banerjee, S., Kroupa, P. \& Oh, S. 2012, ApJ, 746, 15

\bibitem[2008]{bibby}
Bibby, J. L., Crowther, P. A., Furness, J. P. \& Clark, J. S. 2008, MNRAS, 386, L23



\bibitem[1961]{blaauw}
Blaauw, A. 1961,  Bull. Astron. Inst. Netherlands, 15, 265

\bibitem[1999]{bohannan}
Bohannan, B. \& Crowther, P. A. 1999, ApJ, 511, 374

\bibitem[2007]{bonanos}
Bonanos, A. Z. 2007, AJ, 133, 2696

\bibitem[1995]{braun}
Braun, H. \& Langer , N., 1995, A\&A, 297, 483


\bibitem[2011]{brott}
Brott, I., de Mink, S. E., Cantiello, M., et al. 2011, A\&A, 530, 115

\bibitem[2001]{brown}
Brown, G. E., Heger, A., Langer, N., et al. 2001, NewA, 6, 457

\bibitem[1989]{cardelli}
Cardelli, J. A., Clayton, G. C. \& Mathis, J. S. 1989,ApJ, 345, 245

\bibitem[2012]{chini}
Chini, R., Hoffmeister, V. H., Nasseri, A., Stahl, O. \& Zinnecker, H. 2012, MNRAS,
424, 1925

\bibitem[2002]{clark02}
Clark, J. S., Goodwin, S. P., Crowther, P. A., et al., 2002, A\&A, 392, 909 

\bibitem[2005]{clark05}
Clark, J. S., Negueruela, I., Crowther, P. A. \& Goodwin, S. P. 2005, A\&A, 434, 949


\bibitem[2008]{clark08}
Clark, J. S., Muno, M. P., Negueruela, I., et al. 2008, A\&A, 347, 147

\bibitem[2009]{clark09}
Clark, J. S., Davies, B., Najarro, F., et al. 2009, A\&A, 504, 429


\bibitem[2009]{clark09b}
Clark, J. S., Crowther, P. A., Larionov, V. M., et a.. 2009, A\&A, 507, 1555

\bibitem[2010]{clark10}
Clark, J. S., Ritchie, B. W. \& Negueruela, I. 2010, A\&A, 514, 87

\bibitem[2011]{clark11}
Clark, J. S., Ritchie, B. W., Negueruela, I., et al. 2011, A\&A, 531, A28

\bibitem[2012]{clark12}
Clark, J. S., Najarro, F., Negueruela, I., et al. 2012, A\&A, 541, A145

\bibitem[2013]{w9}
Clark, J. S., Ritchie, B. W. \& Negueruela, I. 2013, A\&A,  560, A11

\bibitem[in prep.]{clark13}
Clark, J. S., Goodwin, S. P., Ritchie, B. W. \& Negueruela, I. 2014, A\&A, in prep.

\bibitem[2012]{cottaar}
Cottaar, M., Meyer, M., Andersen, M., Espinoza, P. 2012, A\&A, 539, A5


\bibitem[2006a]{crowther}
Crowther, P. A., Hadfield, L. J., Clark, J. S., Negueruela, I. \& Vacca, W. 
D. 2006a, MNRAS, 372, 1407

\bibitem[2006b]{pacBSG}
Crowther, P. A., Lennon, D. J. \& Walborn, N. R. 2006b, A\&A, 446, 279

\bibitem[1974]{cruz}
Cruz-Gonz\'{a}lez, C., Recillas-Cuz, E., Costero, R., Peimbert, M. \& Torres-Peimbert, S. 1974,
RMxAA,  1, 211

\bibitem[2009]{davies}
Davies, B., Figer, D. F., Kudritzki, R.-P., et al. 2009, ApJ, 707, 844

\bibitem[2006]{dv}
Della Valle,  M. 2006, in Gamma Ray Burts in the Swift Era, Washington, D. C., ed.
S. Holt, N. Gehrels \& J. A. Nousek, AIP, 836, 367

\bibitem[2012]{deller}
Deller, A. T., Camilo, F., Reynolds, J. E. \&  Halpern, J. P. 2012, ApJ, 
748, L1

\bibitem[2003]{dessart}
Dessart, L., Langer, N., Petrovic, J. 2003, A\&A, 404, 991

\bibitem[2010]{dougherty}
Dougherty, S. M., Clark, J. S., Negueruela, I., Johnson, T. \& Chapman, J. M. 2010, A\&A, 511, 58

\bibitem[1992]{duncan}
Duncan, R. C. \& Thompson, C., 1992, ApJ, 392, L9

\bibitem[2012]{ekstrom}
Ekstr\"{o}m, S., Georgy, C., Eggenberger, P. et al. 2012, A\&A, 537, A146

\bibitem[2013]{eldridge}
Eldridge, J. J., Fraser, M., Smartt, S. J., Maund, J. R. \& Crockett, R. M., 2013, MNRAS, 436, 774

\bibitem[2005]{figer05}
Figer, D. F., Najarro, F., Geballe, T. R., Blum R. D. \& Kudritzki, R.P.
2005, ApJ,622, L49

\bibitem[2002]{fryer}
Fryer, C. L., Heger, A., Langer, N. \& Wellstein, S. 2002, ApJ, 578, 335

\bibitem[2012]{fujii}
Fujii, M. S., Saitoh, T. R. \& Portegies Zwart, S. F. 2012, ApJ, 753, 85

\bibitem[2006]{fullerton}
Fullerton, A. W., Massa, D. L. \& Prinja, R. K. 2006, ApJ, 637, 1025



\bibitem[2005]{gaensler}
Gaensler, B. M., McClure-Griffiths, N. M., Oey, M . S., et al. 2005, ApJ, 620, L95


\bibitem[2012]{galyam}
Gal-Yam, A., 2012, Science, 337, 927

\bibitem[2011]{graefener}
Gr\"{a}fener, G., Vink, J. S., de Koter \&  A., Langer, N. 2011, A\&A, 535, 56

\bibitem[2009a]{groh}
Groh, J. H., Hillier, D. J., Damineli, A., et al. 2009a, ApJ, 698, 1698
\bibitem[2009b]{grohb}
Groh, J. H., Hillier, D. J., Damineli, A., et al. 2009b, ApJ, 705, L25
\bibitem[2013]{grunhut}
Grunhut, J. H., Wade, G. A., Leutenegger, M. et al. 2013, MNRAS, 428, 1686

\bibitem[2005]{heger}
Heger, A., Woosley, S. E. \& Spruit, H. C., 2005, ApJ, 626, 350

\bibitem[2007]{helfand}
Helfand, D. J., Chatterjee, S., Brisken, W. F., et al. 2007, ApJ, 662, 
1198 

\bibitem[1998]{hillier98}
Hillier, D. J. \& Miller, D. L., 1998, ApJ, 496, 407
\bibitem[1998b]{hillier98b}
Hillier, D. J., Crowther, P. A., Najarro, F. \& Fullerton, A. W. 
1998, A\&A, 340,  483
\bibitem[1999]{hillier99}
Hillier, D. J., \& Miller, D. L., 1999, ApJ, 519, 354
\bibitem[2001]{hillier01}
Hillier, D. J., Davidson, K., Ishibashi, K. \& Gull, T. 2001, ApJ, 553, 837

\bibitem[2005]{hobbs}
Hobbs, G., Lorimer, D. R., Lyne, A. G. \& Kramer, M. 2005, MNRAS, 360, 974
\bibitem[2013]{inserra}
Inserra, C., Smartt, S. J., Jerkstrand, A., et al. 2013, ApJ, 770, 128
\bibitem[2006]{kaper}
Kaper, L., van der Mer, A. \& Najarro, F., 2006, A\&A, 457, 595

\bibitem[2010]{kasen}
Kasen, D. \& Bildsten, L. 2010, ApJ, 717, 245

\bibitem[2012]{kiminki}
Kiminki, D. C. \& Kobulnicky, H. A. 2012, ApJ, 751, 4

\bibitem[2007]{kothes}
Kothes, R. \&  Dougherty, S. M. 2007, A\&A, 468, 993

\bibitem[2012]{koumpia}
Koumpia, E. \& Bonanos, A. Z. 2012, 547, A30

\bibitem[1994]{kouv}
Kouveliotou, C., Fishman, G. J., Meegan, C. A., et al. 1994, Nature, 368, 125

\bibitem[2012]{ku}
Kudryavtseva, N., Brandner, W., Gennaro, M., et al. 2012, ApJ, 750, L44

\bibitem[2013]{kun}
Kuncarayakti, H., Doi, M., Aldering, G. et al. 2013, AJ, 146, 30

\bibitem[2003]{laming}
Laming, J. M.  \& Hwang, U., 2003, ApJ, 597, 347
\bibitem[1989]{langer89}
Langer, N., 1998, A\&A, 210, 93
\bibitem[1998]{langer}
Langer, N., 1998, A\&A, 329, 551
\bibitem[2012]{ARAA}
 Langer, N., 2012, ARAA, 50, 107
\bibitem[2007]{lefever}
Lefever, K., Puls, J. \& Aerts, C. 2007, A\&A, 463, 1093
\bibitem[2008]{linder}
Linder, N., Rauw, G., Martins, F., et al. 2008, A\&A, 489, 713

\bibitem[2009]{luna}
Luna, A., Mayya, Y. D., Carrasco, L., Rodr\'{i}guez-Merino, L. H., \&  Bronfman, L. 
2009, RMxAC, 37, 32

\bibitem[2000]{maeder}
Maeder, A. \& Meynet, G. 2000, A\&A, 361, 159
\bibitem[2008]{markova}
Markova, N. \& Puls, J. 2008, A\&A, 478, 822
\bibitem[2007]{martinsGC}
Martins, F., Genzel, R., Hillier, D. J., et al. 2007, A\&A, 468, 233
\bibitem[2012]{mason}
Mason, A. B., Clark, J. S., Norton, A. J., et al. 2012, A\&A, 422, 199 
\bibitem[2013]{mccrum}
McCrum, M., Smartt, S. J., Kotak, R. et al. 2013, MNRAS, 437, 656
\bibitem[2009]{mengel}
Mengel, S. \& Tacconi-Garman, L. E. 2009, ApSS, 324, 321
\bibitem[2008]{mereghetti}
Mereghetti, S., A\&ARv, 15, 225
\bibitem[2011]{metzger}
Metzger, B. D., Giannios, D., Thompson, T. A., Bucciantini, N., Quataert, E., 2011, MNRAS,
413, 2031
\bibitem[2000]{meynet}
Meynet, G. \& Maeder, A. 2000, A\&A,, 361, 101
\bibitem[2007]{mokiem}
Mokiem, M. R., de Koter, A., Vink, J. S., et al. 2007, A\&A, 473, 603
\bibitem[2013]{mori}
Mori, K., Gotthelf, E. V., Zhang, S. et al. 2013, ApJ, 770, L23
\bibitem[2007]{muno}
Muno, M.P. 2007, AIPC, 924, 166
\bibitem[2006a]{muno06a}
Muno, M. P., Clark, J. S., Crowther, P. A., et al.  2006a, ApJ, 636 L41
\bibitem[2006b]{muno06b}
Muno, M. P., Law, C., Clark, J. S., et al. 2006, ApJ, 650, 203
\bibitem[2008]{muno08}
Muno, M. P., Gaensler, B. M., Nechita, A., Miller, J. M. \& Slane, P. O. 2008, ApJ, 680, 639 
\bibitem[1997]{paco97}
Najarro, F., Hillier, D. J. \& Stahl, O. 1997, A\&A, 326, 1117
\bibitem[2001]{paco01}
Najarro, F. 2001, ASP Conf. Ser. 233, P Cygni 2000: 400 years of Progress, ed.
M. de Groot \& C. Sterken (San Francisco, CA:ASP), 133
\bibitem[2006]{paco06}
Najarro, F., Hillier, D. J., Puls, J., Lanz, T. \& Martins, F. 2006, A\&A,
456, 659
\bibitem[2009]{paco09}
Najarro, F., Figer, D. F., Hillier, D. J., Geballe, T. R. \& Kudritzki, R. P. 2009, ApJ, 691, 1816
\bibitem[2010]{iggy10}
Negueruela, I., Clark,
 J. S. \& Ritchie, B. W. 2010, A\&A, 516, A78
\bibitem[2013]{nicholl}
Nicholl, M., Smartt, S. J., Jerkstrand, A., 2013, Nature, 503, 236
\bibitem[1982]{nomoto}
Nomoto, K., Sugmioto, D., Sparks, W. M., et al. 1982, Nature, 299, 803 
\bibitem[2002]{pasquini}
Pasquini, L., Avila, G., Blecha, A., et al. 2002, The Messenger, 110, 1
\bibitem[2005]{petrovic}
Petrovic, J., Langer, N. \& van der Hucht, K. A., 2005, A\&A, 435, 1013

\bibitem[1967]{poveda}
Poveda, A., Ruiz, J. \& Allen, C. 1967, Bol. Obs. Tonantzintla y
Tacubaya, 4, 86 

\bibitem[2011]{quimby}
Quimby, R. M., Kulkarni, S. R., Kasliwal, M. M., et al. 2011, Naturem 474, 487


\bibitem[2012]{reig}
Reig, P., Torj\'{o}n, J. M. \&  Blay, P., 2012, MNRAS, 401, 595


\bibitem[2009a]{ben09}
Ritchie, B. W., Clark, J. S., Negueruela, I. \& Crowther, P. A. 2009a, A\&A, 507, 1585

\bibitem[2009a]{ben09b}
Ritchie, B. W., Clark, J. S., Negueruela, I. \&  Najarro, F. 2009b, A\&A,
507, 1597

\bibitem[2010]{ben10}
Ritchie, B. W., Clark, J. S., Negueruela, I. \& Langer, N. 2010, A\&A, 520, A48

\bibitem[in prep.]{ben12}
Ritchie, B. W., Clark, J. S. \& Negueruela, I. 2014, A\&A, in prep. 


\bibitem[2012]{sana12}
Sana, H., de Mink, S. E., de Koter, A., et al., 2012, Science, 337, 444


\bibitem[2008]{searle}
Searle, S. C., Prinja, R. K., Massa, D. \& Ryans, R. 2008, A\&A, 481, 777

\bibitem[2011]{smith}
Smith, N. Li,W., Filippenko, A. V., \& Chornock, R. 2011, MNRAS, 412, 1522

\bibitem[2008]{spruit}
Spruit, H. C. 2008, in 40 years of pulsars:millisecond pulsars, magnetars and more. AIP Conference proceedings, Vol. 983, p. 391

\bibitem[2012]{tendulkar}
Tendulkar, S. P., Cameron, P. B. \& Kulkarni, S. R. 2012, ApJ, 761, 761
\bibitem[2013]{tendulkar13}
Tendulkar, S. P., Cameron, P. B. \& Kulkarni, S. R. 2013, ApJ, 772, 31
\bibitem[2011]{tetzlaff}
Tetzlaff, N., Neuh\"auser, R. \& Hohle, M. M. 2011, MNRAS, 410 190

\bibitem[1993]{thompson}
Thompson, C. \& Duncan, R. C. 1993, ApJ, 408, 194
\bibitem[2004]{thompson04}
Thompson, T. A., Chang, P. \& Quataert, E. 2004, ApJ, 611, 380
\bibitem[2008]{tout}
Tout, C. A., Wickramasinghe, D. T., Liebert, J., Ferrario, L. \& Pringle, J. E. 2008, MNRAS, 387, 897 
\bibitem[1992]{usov}
Usov, V. V. 1992, Nature, 357, 472
\bibitem[2000]{vink00}
Vink, J. S., de Koter, A. \& Lamers, H. J. G. L. M. 2000, A\&A, 362, 295
\bibitem[2006]{vink}
Vink, J. \& Kuiper, L., 2006, MNRAS, 370, L14
\bibitem[1965]{vitrichenko}
Vitrichenko, E. A., Gershberg, R. E. \& Metik, L. P. 1965, IzKry, 34, 193  
\bibitem[1999]{wellstein}
Wellstein, S. \& Langer, N. 1999, A\&A, 350, 148
\bibitem[2001]{wellstein01}
Wellstein, S., Langer, N. \& Braun, H., 2001, A\&A, 369, 939
\bibitem[2006]{woods}
Woods, P. M. \& Thompson, C. 2006 in Compact stellar X-ray sources. Edited by W. Lewin \& M. van der Klis. Cambridge Astrophysics
Series, No. 39, Cambridge University Press, p. 547
\bibitem[2008]{woods08}
Woods, P. M., 2008, in 40 years of pulsars:millisecond pulsars, magnetars and more. AIP Conference proceedings, Vol. 983, p. 227
\bibitem[2011]{woods11}
Woods, P. M., Kaspi, V. M., Gavriil, F. P. \& Airhart, C. 2011, ApJ, 726, 37
\bibitem[2010]{woosley10}
Woosley, S. E.  2010, ApJ, 719, L204
\bibitem[2010]{yoon}
Yoon, S.-C., Woosley, S. E. \& Langer, N. 2010, ApJ, 725, 940
\bibitem[2006]{young}
Young, P. A., Fryer, C. L., Hungerford, A., et al. 2006, ApJ, 640, 891
\end{thebibliography}
\end{document}